\pdfoutput=1
\documentclass[aps,prl,reprint,amsmath,amssymb,superscriptaddress]{revtex4-1}
\usepackage{bm}
\usepackage{graphicx}
\usepackage[colorlinks]{hyperref}
\usepackage{mathrsfs}
\usepackage{times}
\usepackage[normalem]{ulem}

\newcommand{\beq}{\begin{equation}}
\newcommand{\eeq}{\end{equation}}
\newcommand{\hpa}{h_{\parallel}}
\newcommand{\hpe}{h_{\perp}}

\newcommand{\dd}{{\rm d}}
\newcommand{\ff}{\mathfrak{f}}

\begin{document}

\newcommand{\titleinfo}{
Fragmentation and emergent integrable transport in the weakly tilted Ising chain
}

\title{\titleinfo}

\author{Alvise Bastianello}
\affiliation{Department of Physics, Technical University of Munich, 85748 Garching, Germany}
\affiliation{Institute for Advanced Study,  85748 Garching, Germany}
\affiliation{Munich Center for Quantum Science and Technology (MCQST), Schellingstr. 4, D-80799 M{\"u}nchen, Germany}

\author{Umberto Borla}
\affiliation{Department of Physics, Technical University of Munich, 85748 Garching, Germany}
\affiliation{Munich Center for Quantum Science and Technology (MCQST), Schellingstr. 4, D-80799 M{\"u}nchen, Germany}

\author{Sergej Moroz}
\affiliation{Department of Physics, Technical University of Munich, 85748 Garching, Germany}
\affiliation{Munich Center for Quantum Science and Technology (MCQST), Schellingstr. 4, D-80799 M{\"u}nchen, Germany}
\affiliation{Department of Engineering and Physics, Karlstad University, Karlstad, Sweden}

\begin{abstract}
We investigate emergent quantum dynamics of the tilted Ising chain in the regime of a weak transverse field.
Within the leading order perturbation theory, the Hilbert space is fragmented into exponentially many decoupled sectors.
We find that the sector made of isolated magnons is integrable with dynamics being governed by a constrained version of the XXZ spin Hamiltonian. 
As a consequence, when initiated in this sector, the Ising chain exhibits ballistic transport on unexpectedly long times scales. We quantitatively describe its rich phenomenology employing exact integrable techniques such as Generalized Hydrodynamics.
Finally, we initiate studies of integrability-breaking magnon clusters whose leading-order transport is activated by scattering with surrounding isolated magnons.

\end{abstract}

\maketitle

%%%%%%%%%%%%%%%%%%%%%%%%%%%%%%%%%
%%%%%%%%%%%%%%%%%%%%%%%%%%%%%%%%%

\paragraph{Introduction.---}
The celebrated Ising model contributed to several paradigm shifts in physics. In classical statistical mechanics, Onsager's solution \cite{onsager1944crystal} on a two-dimensional lattice kick-started the development of a general theory of continuous phase transitions. 
It is well-known that in the presence of a transverse field only, the one-dimensional quantum Ising chain is exactly solvable \cite{PhysRevLett.106.227203,Calabrese_2012,Calabrese2_2012} and reduces via the Jordan-Wigner transformation to a free Majorana chain \cite{schultz1964two}. On the other hand, the addition of a longitudinal field breaks the integrability of the model. In the ferromagnetic case, this leads to confinement of fermionic domain walls into bosonic magnons \cite{PhysRevD.18.1259,DELFINO1996469,Rutkevich2008,Fonseca2003}. Recent studies concentrated on aspects of anomalously slow dynamics \cite{kormos2017real, PhysRevB.99.180302, PhysRevB.102.014308, PhysRevB.102.041118, vovrosh2020confinement,PhysRevLett.124.230601}, quantum scarring \cite{PhysRevLett.122.130603, PhysRevB.99.195108}, prethermalization \cite{de2019very}, fractons \cite{PhysRevResearch.2.013094}, meson scattering \cite{surace2021scattering, karpov2020spatiotemporal}, dynamics of the false vacuum \cite{Sinha_2021,lagnese2021false,milsted2021collisions,rigobello2021entanglement,Javier_Valencia_Tortora_2020,pomponio2021bloch} and emergent $\mathbb{Z}_2$ lattice gauge theories \cite{PhysRevB.99.075103, borla2021gauging}.

\begin{figure}[t!]
	\includegraphics[width=1\columnwidth]{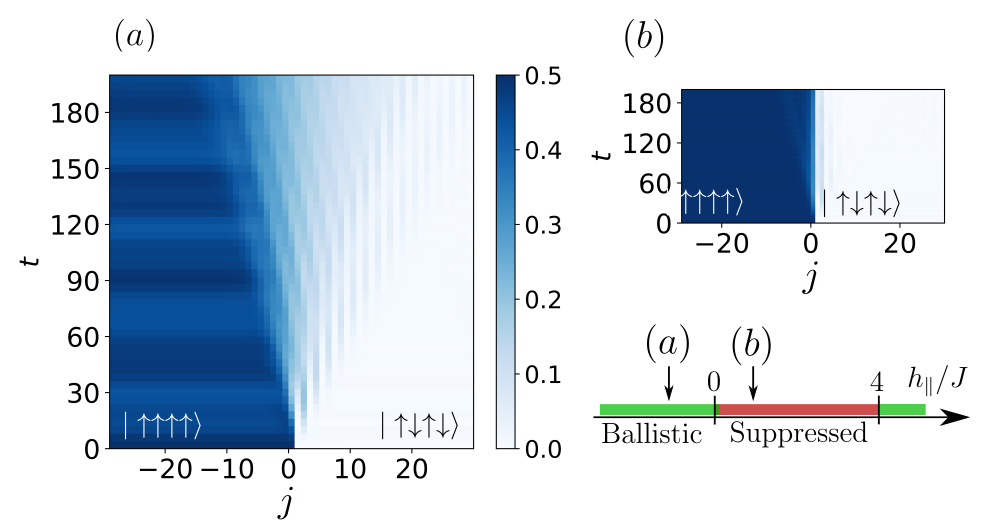}
	\caption{
	Magnetization profiles $\langle S^z_j\rangle$ in the Ising chain at $J=1$ and $\hpe=0.2$ initialized by joining the ferromagnetic and Neel states. 
	(\textit{a}) For $\hpa=-0.7$  we observe ballistic transport with a characteristic lightcone. (\textit{b}) For $\hpa=0.7$ we find strong suppression of spin transport.
	 TEBD simulations are done for a chain of length $L=80$.
	The peculiar transport is captured by the integrable dynamics governed by the Hamiltonian \eqref{ABform} which emerges for a weak transverse field $\hpe\ll (\hpa,J)$. The validity of the phase diagram is within this limit, see main text for discussion.
	}
	\label{fig_partitioning}
\end{figure}

In this paper, we unveil unexpected features of the one-dimensional quantum Ising model in a weakly tilted field.
Specifically, we investigate transport in the prototypical partitioning protocol \cite{Bernard_2016}: the chain is initialized into two halves which are then connected, activating transport across the junction.  We observe strong signatures of ballistic behavior for unexpectedly long times in the regime where the transverse field is small. Moreover, we discover that the nature of transport exhibits a strong dependence on the longitudinal field and on the Ising coupling.
Using degenerate perturbation theory as a tool, we argue that the effective Hamiltonian in this regime enjoys two separate $U(1)$ conservation laws for the number of magnons and domain walls. These two symmetries are emergent as they are not imprinted in the microscopic Hamiltonian. We show that, at leading order in perturbation theory, the effective dynamics fragments the Hilbert space (expressed in the canonical local basis) into a large number of independent sectors that scales exponentially in the system size. Among all sectors we first zoom in on the dynamics of isolated magnons, which we find to be integrable.
This finding accounts for the emergence of ballistic behavior -- a clear signature of integrability-- in contrast to the naively-expected diffusion.
Specifically, this sector is governed by the constrained XXZ Hamiltonian first investigated by Alcaraz and Bariev \cite{Alcaraz_Bariev} with coordinate Bethe Ansatz.
Apart from early studies \cite{karnaukhov2002one, alcaraz2007exactly} this model went unnoticed for a long time, but recently appeared in several independent contexts, e.g. in the constrained PXXP model \cite{verresen2019stable}, in the strongly-coupled regime of a $\mathbb{Z}_2$ lattice gauge theory coupled to fermions \cite{PhysRevLett.124.120503} and in interacting correlated hopping models \cite{PhysRevLett.124.207602}.
At the non-interacting point, that is non-trivial due to the constraint, 
%it has been shown to emerge 
it emerges in the strong coupling limit of the canonical XXZ spin chain \cite{zadnik2020folded, zadnik2020folded2, pozsgay2021integrable}. 
The leitmotif of some of these studies is the phenomenon of Hilbert space fragmentation \cite{PhysRevX.10.011047, PhysRevB.101.174204, moudgalya2019thermalization, papic2021weak, Moudgalya:2021xlu} due to imposed or emergent constraints which make the constrained XXZ chain a natural candidate to describe integrable sectors, if present.
Moreover, see also Refs. \cite{pozsgay2021yangbaxter, pozsgay2021integrable1, gombor2021integrable} for related integrable constrained models with medium range interactions.
Using Generalized Hydrodynamics (GHD) \cite{PhysRevX.6.041065,PhysRevLett.117.207201} (see also Refs. \cite{Bastianello_2022,denardis2021correlation,alba2021generalizedhydrodynamic,bulchandani2021superdiffusion,bastianello2021hydrodynamics,borsi2021current,cubero2021form}) we analytically tackle transport within the isolated magnon sector.
The Alcaraz-Bariev (AB) model inherits the rich phenomenology of the XXZ spin chain:
transport greatly depends on the interactions and can exhibit sharp jumps \cite{PhysRevB.96.115124}.
We find that the hydrodynamics of the AB model is peculiar on its own, since in certain regimes of interactions quasiparticles carry fractional magnetization, in clear contrast with the vast majority of integrable models and signaling the collective nature of the excitations.
The presence of two or more neighboring magnons breaks integrability and probes the transport of surrounding isolated magnons.
Indeed, within the leading order perturbation theory clusters of magnons are completely immobile in isolation, but we show they undergo magnon-assisted hopping experiencing biased diffusion, whose mean and variance are directly connected to the magnetization current crossing them.
%%%%%%%%%%%%%%%%%
\paragraph{Emergent ballistic transport in the Ising chain in a weak transverse field---} With the help of Time Evolving Block Decimation (TEBD) \cite{SCHOLLWOCK201196}, we start by numerically investigating transport in 
the Ising chain in a tilted magnetic field
\beq \label{HIsing}
\begin{split}
H&= -J \sum_{i} Z_{i} Z_{i+1}- \hpa \sum_i Z_i - \hpe \sum_i X_i,
\end{split}
\eeq
where $X_i$ and $Z_i$ denote the Pauli matrices at site $i$.
In the partitioning protocol \cite{Bernard_2016}, one initializes the state in two different halves $|\Psi\rangle=|\Psi_L\rangle\otimes |\Psi_R\rangle$ and then lets the system evolve with the homogeneous Hamiltonian. 
In Fig. \ref{fig_partitioning} $(a)$ we choose $|\Psi_L\rangle$ and $|\Psi_R\rangle$ to be the Neel and ferromagnetic state respectively, and we focus on the regime where the transverse field is weak.
While the Hamiltonian \eqref{HIsing} is known to be non-integrable for generic values of the parameters, our analysis unveils persistent ballistic transport typical of integrable models \cite{PhysRevX.6.041065,PhysRevLett.117.207201}, in contrast with the naively expected diffusion.
With this choice of initial states, we also observe a strong dependence of transport on the longitudinal field and the Ising coupling with a lightcone suppression whenever $0<\hpa/J<4$, see Fig. \ref{fig_partitioning} $(b)$.
This unexpected behavior can be ascribed to a peculiar integrable model, as we now discuss.

%%%%%%%%%%%%%%%%%%%%%%%%%%%
\paragraph{Effective Hamiltonian.---}
We analyse the Ising chain \eqref{HIsing} in the regime where the transverse field $\hpe$ is much smaller than the two generic (but incommensurate) couplings $J$ and $\hpa$.  To set up a perturbative expansion we split the Hamiltonian \eqref{HIsing} into the classical $Z$-dependent part $H_0$ (the Ising and longitudinal field terms) and  the transverse field perturbation. Since $[H_0, Z_i]=0$, the Hamiltonian $H_0$ has an extensive number of symmetries and trivially splits in the $Z$-basis into $2^L$ independent blocks. Notwithstanding, its energy spectrum is organized into degenerate multiplets characterized  only by a pair of emergent $U(1)$ quantum charges: the magnon number $N$ and the domain wall number $D=\sum_i (1-Z_i Z_{i+1})/2$. 
By construction, $N$ and $D$ are both simultaneously preserved by the effective perturbative dynamics.
The transverse field perturbation changes the number of magnons and thus can contribute only at even orders of the degenerate perturbation theory. Employing the Schrieffer-Wolff transformation \cite{PhysRev.149.491, bravyi2011schrieffer}, in the Supplementary Material (SM) \footnote{Supplementary Material for construction of the effective Hamiltonian; integrability and hydrodynamics of the Alcaraz-Bariev model; analysis of the energy level statistics; dynamics of a two-magnon cluster.} we have constructed the second-order effective perturbative Hamiltonian
\beq \label{eff2}
\begin{split}
H^{(2)}_{eff}=&-\sum_{s=\pm 1} t_s  \sum_j \mathcal{P}^{s}_{j-1, j+2} \left( S^+_j S^-_{j+1}+h.c.  \right) \\
& -g  \sum_j Z_{j-1}Z_{j}Z_{j+1} -\delta J  \sum_j Z_{j}Z_{j+1} -\delta \hpa  \sum_j Z_{j},
\end{split}
\eeq
where the spin-exchange coupling $t_s=\hpe^2 \hpa^{-1} J/(\hpa+2sJ)$, the projector $\mathcal{P}^{s}_{i, j}=\left(1+s (Z_i +Z_j)+ Z_i Z_j \right)/4$ and $S^\pm_j=(X_j \pm i Y_j)/2$. Moreover, the induced three-spin coupling $g=\hpe^2 \hpa^{-1} J^2/\alpha$ and the shifts of the Ising and longitudinal couplings are $\delta J=-\hpe^2 J/ \alpha$ and $\delta \hpa=\hpe^2 \hpa^{-1} (\hpa^2-2 J^2)/(2\alpha)$, where we introduced $\alpha=\hpa^2-4 J^2$. 
Corrections beyond Eq. \eqref{eff2} are $ \mathcal{O}(\hpe^4)$ and are discussed in \cite{Note1}.
The Hamiltonian agrees with the previous derivation \cite{PhysRevA.95.023621}, see also \cite{PhysRevLett.124.207602,karpov2020spatiotemporal} for related studies.
Domain wall conservation enforces the projector $\mathcal{P}^{s}_{i, j}$ ensuring that the two outer spins surrounding the exchange pair point in the same direction. 
Similar type of hopping have been recently discussed in  \cite{PhysRevLett.124.207602, zadnik2020folded, zadnik2020folded2, pozsgay2021integrable}. Since only isolated magnons can hop, the perturbative model \eqref{eff2} supports a large number of immobile (frozen) quantum states that contain clusters of magnons. 

The number $F_l$ of independent frozen states of size $l\gg1$ scales exponentially $F_l\sim\varphi^l$, where $\varphi$ is the golden ratio \cite{PhysRevLett.124.207602}.
In SM \cite{Note1} we demonstrate that for chains of size $L\gg 1$ the effective Hamiltonian \eqref{eff2} splits into $\varphi^{L+1}$ independent blocks. Such exponential growth is parametrically larger than the $O(L^2)$ scaling expected purely from the two $U(1)$ emergent symmetries. A similar pattern of fragmentation of the Hilbert space was discovered in spin models in the strict confinement regime \cite{PhysRevLett.124.207602}.
Consider first a sector with $N$ isolated spin-down sites in the spin-up background. In this case $D=2N$ and pairs of magnons cannot appear next to each other. In this sector the second-order Hamiltonian \eqref{eff2} reduces to
\beq \label{ABform}
H_{eff}^{(2)}\to- \mathcal{J} \sum_j \mathcal{P}_{1} \left( S^x_j S^x_{j+1}+S^y_j S^y_{j+1}+ \Delta S_j^z S_{j+2}^z   \right) 
\mathcal{P}_{1},
\eeq
where the projector $\mathcal{P}_{1}$ prohibits two spin-down magnons to occupy neighbouring sites.
The inverse of the coupling $\mathcal{J}=2t_+$ defines the slow time scale associated with hopping of the isolated magnons. The anisotropy parameter $\Delta=2J/(\hpa -2J)$ can be tuned by changing the dimensionless ratio $\hpa/J$.
This model is a constrained version of the celebrated XXZ chain which was first investigated by Alcaraz and Bariev \cite{Alcaraz_Bariev}. 
Remarkably, the Hamiltonian \eqref{ABform} at $\Delta=1/2$ is known to be a supersymmetric model \cite{PhysRevLett.90.120402,Fendley_2003}, which can be realized in a Rydberg-based quantum simulator \cite{minar2020kink}.

%%%%%%%%%%%%%%%%%%%%%%%%%%%%%%%%%%%%%%%%%%%%%%%%%

\paragraph{Transport in the Alcaraz-Bariev model.---} 
The Alcaraz-Bariev (AB) model can be generalized to the extended hard-core constraint $\mathcal{P}_1\to \mathcal{P}_T$ prohibiting magnons closer than $T$ sites.
The original papers \cite{Alcaraz_Bariev,karnaukhov2002one, alcaraz2007exactly} addressed the equilibrium thermodynamics. 
For $\Delta=0$ and $T=1$, the AB model  governs the isolated magnon sector of the folded XXZ spin chain \cite{zadnik2020folded, zadnik2020folded2, pozsgay2021integrable}. Here we focus on transport and hydrodynamics of the AB model at arbitrary $\Delta$.

Being integrable, the AB model possesses an extensive number of (quasi-)local conserved quantities \cite{Ilievski_2016}, with striking consequences on its nonequilibrium features, hindering thermalization \cite{PhysRevLett.98.050405} and featuring ballistic transport \cite{denardis2021correlation}.
The AB Hilbert space is made of multiparticle magnonic asymptotic states labeled by the set of rapidities $\{\lambda_j\}_{j=1}^N$, which generalize the momenta of non-interacting systems. 
Due to integrability, multiparticle scattering events can be factorized in two-body scattering processes, the latter fully described by the scattering phase $\Theta(\lambda,\lambda')$.
The scattering phase of the AB model and of the XXZ spin chain are connected \cite{Alcaraz_Bariev,Note1} $\Theta(\lambda,\lambda')=Tp(\lambda)-Tp(\lambda')+\Theta^{XXZ}(\lambda-\lambda')$, with $p(\lambda)$ the momentum of the magnon. The relation resembles the celebrated $T\bar{T}$ deformation, see \cite{zamTT,jiang2020mathrmtoverlinemathrmtdeformed,cardy2021toverline,doyon2021space,PhysRevD.103.066012,PhysRevLett.124.100601,Pozsgay2020} and references therein.
On a finite chain, the allowed rapidities are quantized, similarly to the momenta of non-interacting models. However, the interactions couple the rapidities through the highly non-linear Bethe equations \cite{takahashi2005thermodynamics,Note1}, which explicitly depend on $\Theta$.
Being non-linear, the Bethe equations are difficult to solve. 
In the zero density limit ($L\to\infty$, $N$ fixed), the solutions of the Bethe equations form groups of rapidities sharing the same real part, but shifted in the imaginary direction. These special solutions are called strings and are determined by the zeroes and poles of the scattering matrix $e^{i\Theta(\lambda,\lambda')}$ \cite{takahashi2005thermodynamics} and are readily interpreted  as bound states of magnons. Since the factor $e^{iT(p(\lambda)-p(\lambda'))}$ does not have zeroes or poles, in the AB scattering matrix these are entirely determined by the XXZ scattering matrix. Hence the two models share the same pattern of strings.

The string hypothesis \cite{takahashi2005thermodynamics} claims the persistence of strings even in the thermodynamic limit ($L\to\infty$, $N/L=n$ fixed).
Within the Thermodynamic Bethe Ansatz (TBA) \cite{takahashi2005thermodynamics}, one opts for a coarse-grained description of the Bethe equations, defining the so called root densities $\rho_j(\lambda)$, one for each string, where $\lambda$ parametrizes the (real) center of the string. Then, $L\dd\lambda \rho_j(\lambda)$ is interpreted as the number of solutions of the $j^\text{th}$ string within the interval $[\lambda,\lambda+\dd \lambda]$. 
The interactions affect the occupancy, hence the need of introducing the total root density $\rho_j^t(\lambda) \ge \rho_j(\lambda)$ representing full occupancy (see SM for details \cite{Note1}).
The root densities fully determine the equilibrium thermodynamics and homogeneous nonequilibrium steady states \cite{PhysRevLett.110.257203,Caux_2016,Ilievski_2016}. Moreover, they are the building blocks of GHD.
Since the AB and XXZ models are closely related, it is worth to address properly the string hypothesis in the latter. The string classification in the XXZ chain is textbook material \cite{takahashi2005thermodynamics} and we summarize it in SM \cite{Note1}. The structure of XXZ strings greatly depends on the parameter $\Delta$: in particular, for $|\Delta|\ge 1$ the string hypothesis, strictly speaking, does not cover the entire phase space. The thermodynamics of the strings built on the all-spin-up reference state covers only states up to half filling $0<n<1/2$, with $n$ being the density of flipped spins.
In the XXZ model, one circumvents this limitation by using the reflection symmetry $S^z_j\to -S^z_j$ and building the string hypothesis on the symmetric all-spin-down reference state. The two descriptions together cover the whole phase space and, in addition to the root densities, one introduces the magnetization sign $\ff=\pm 1$ to specify the sector.  In the case $|\Delta|<1$, the string hypothesis covers all magnetization sectors and $\ff$ is not needed.

In the AB model, the constraint shifts the half-filling point to the value $1/(2+T)$. Moreover, it breaks the spin reflection symmetry.
In Ref. \cite{Alcaraz_Bariev} the Bethe equations of the AB model in all sectors have been mapped onto the corresponding equations for the XXZ chain in a reduced magnetization-dependent volume. Building on these ideas, we now determine the thermodynamics of the AB model at a generic filling, which is described by the same set of root densities as the XXZ spin chain.  Above half filling, these cannot be interpreted as strings anymore;
however, for  the sake of retaining a standard notation, we will still refer to these root densities as strings.
In addition, for $|\Delta|>1$ one needs an extra bit of information $\ff=\pm 1$ that distinguishes the regions below and above half filling, respectively. 
When addressing thermodynamics and transport, it is crucial to know the amount of magnetization carried by each string. Within the ordinary string hypothesis, this is simply the number of magnons belonging to the same bound state. In the XXZ case, one has $m_j^{XXZ}=\ff |m_j^{XXZ}|$, with $|m_j^{XXZ}|$ a $\ff-$independent integer. On the other hand, in the AB model we find an explicitly $\ff-$dependent magnetization $m_j=[1+T(1-\ff)/2]^{-1}m_j^{XXZ}$. We observe that for $\ff=-1$ (needed if $|\Delta|>1$) the string magnetization $m_j$ becomes fractional! 
This signals the lack of microscopic interpretation of the root density as a bound state of magnons.
We found that the non-trivial $\ff-$dependence extends from the magnetization to thermodynamic observables. To see that we consider the TBA string scattering phase $\Theta_{j,j'}(\lambda,\lambda')$ that, whenever the string hypothesis holds, is obtained from $\Theta(\lambda,\lambda')$ summing over the constituents of the string. In all sectors it can be written as
\beq
\Theta_{j,j'}(\lambda,\lambda')=T p_j(\lambda) m_{j'}-T m_{j}p_{j'}(\lambda')+\Theta_{j,j'}^{XXZ}(\lambda-\lambda')\, .
\eeq
The appearance of the magnetization $m_j$ makes $\Theta_{j,j'}$ explicitly $\ff-$dependent. In addition, we find that $\ff$ renormalizes the total root density  $2\pi\sigma_j\rho^t_j=(\partial_\lambda p_j)^\text{dr}(1+T(1-\ff)/2)^{-1}$, where $\sigma_j$ is the string parity and the standard definition of dressing is
$(\partial_\lambda p_j)^\text{dr}=\partial_\lambda p_j-\sum_{j'}\int \frac{\dd\lambda}{2\pi}\partial_\lambda \Theta_{j,j'}(\lambda,\lambda')\vartheta_{j'}(\lambda')\sigma_{j'}(\partial_{\lambda'} p_{j'})^\text{dr}$, with $\vartheta_j=\rho_j/\rho_j^t $ being the filling fraction. With these caveats, one can recover the full equilibrium thermodynamics by standard methods: we leave the details to SM \cite{Note1} and move on towards discussing hydrodynamics. 
\begin{figure}[t]
	\includegraphics[width=0.98\linewidth]{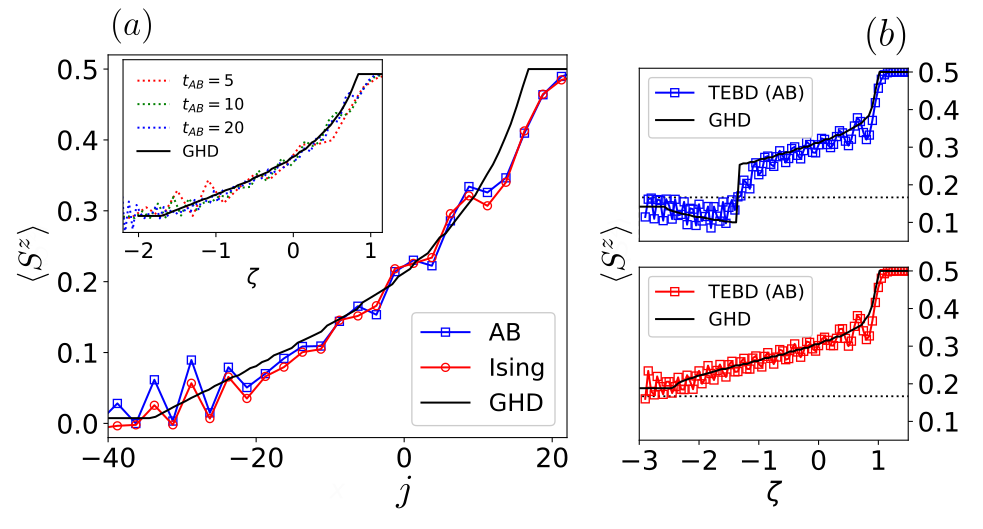}
		\caption{\label{fig_transport}
		(\emph{a}) The magnetization profile of a chain of length $L=80$ evolved with TEBD from $|\text{Neel}\rangle\otimes |\text{ferro}\rangle$ at large time (measured in the AB units $[\mathcal{J}^{-1}]$) $t_{\text{AB}}=20$ approaches the GHD prediction. For the Ising model we choose parameters $\hpe=0.5$, $\hpa=6$ and $J=1$, corresponding to $\Delta = 0.5$ in the AB model. In the inset, we show the collapse of the AB simulations on the GHD analytical prediction.
		 (\emph{b}) To highlight magnetization jumps  in $|\Delta|>1$ (precisely, $\Delta=1.5, \mathcal{J}=-1$), we consider the partitioning from $|\text{GS}_{\langle Z\rangle}\rangle\otimes|\text{ferro}\rangle$ with $|\text{GS}_{\langle Z\rangle}\rangle$ the ground state of the AB model in the sector at fixed magnetization $\langle Z\rangle$ for a chain of length $L=120$. For the left-side magnetization being below (top) and above (bottom) the half-filling dotted line, the profile exhibits qualitatively different behaviour.
		 }
\end{figure}
Let us imagine that the system, still governed by the homogeneous AB Hamiltonian, features a long wavelength inhomogeneity in the state. In the limit of weak inhomogeneities, one can invoke local relaxation to  (weakly) space-time dependent root densities. This is the idea behind GHD \cite{PhysRevX.6.041065,PhysRevLett.117.207201}, which in its simplest form describes the convective expansion of particles $\partial_t \rho_j(\lambda)+\partial_x[v^\text{eff}_j(\lambda)\rho_j(\lambda)]=0$. The effective velocity 
\beq\label{eq_veff}
v^\text{eff}_j(\lambda)=(\partial_\lambda \epsilon_j(\lambda))^\text{dr}/(2\pi \sigma_j \rho^t_j(\lambda))\, ,
\eeq 
depends on the state due to interactions, making the equation non-linear. Above, $\epsilon_j$ is the energy carried by the string.
In contrast to the AB model, in most integrable systems the identity $2\pi\sigma_j\rho^t_j=(\partial_\lambda p_j)^\text{dr}$ holds, leading to the alternative more intuitive definition $v^\text{eff}_j(\lambda)=(\partial_\lambda \epsilon_j)^\text{dr}/(\partial_\lambda p_j)^\text{dr}$ that was originally reported in Refs. \cite{PhysRevX.6.041065,PhysRevLett.117.207201}. However, in a recent rigorous proof \cite{
PhysRevX.10.011054,PhysRevLett.125.070602,borsi2021current}, Eq. \eqref{eq_veff} naturally emerges from the calculations.  
At a technical level, Eq. \eqref{eq_veff} arises in the AB model naturally by manipulating the hydrodynamic equations \cite{Note1}.
To the extent of our knowledge, this is the only model with this feature.
In the case with $|\Delta|>1$, the spin flip continuity  $\partial_t n+\partial_x j_n=0$, with $n=(1-\ff)/(2+T(1-\ff))^{-1}+\sum_j\int \dd \lambda m_j\rho_j(\lambda)$ and $j_n=\sum_j\int \dd \lambda v^\text{eff}_j(\lambda)m_j\rho_j(\lambda)$, closes the hydrodynamic equations giving a further condition on $\ff$, similarly to the XXZ model \cite{PhysRevB.96.115124}.

\paragraph{The partitioning protocol and GHD.---} 
We now apply GHD of the AB model to the partitioning protocol.
After a short transient the profile of local observables becomes scale-invariant \cite{PhysRevX.6.041065,PhysRevLett.117.207201} $\langle \mathcal{O}(t,x)\rangle=F[x/t]$ and curves at different time collapse when plotted as a function of the ray $\zeta=x/t$.
As we show in Fig. \ref{fig_transport} $(a)$, if one starts from an initial state with only isolated magnons the Ising chain agrees with the underlying AB description  (up to a time scale $t\sim \hpe^{-4}$) and supports ballistic transport.
Note that for $|\Delta|\ge 1$, i.e. $0\le\hpa/J\le 4$, the magnetization sign $\ff$ is responsible of sharp jumps whenever states from the two different magnetization sectors are joined.
At $t=0$, the $\ff(x)$ profile is a step function and due to discreteness of $\ff$, GHD cannot smoothen its profile, but only moves the position of the jump. 
The explicit $\ff-$dependence of the TBA induces non-analyticities not only in the magnetization profile (as in the XXZ chain \cite{PhysRevB.96.115124}), but in all conserved charges.

An extreme example is presented in Fig. \ref{fig_partitioning}: for $|\Delta|\ge 1$, the Neel state and the ferromagnetic states have the exactly same trivial root density $\rho_j(\lambda)=0$, but differ in the sign of $\ff$ \cite{Note1}.
Hence, any smooth dependence of the profile is suppressed and only the jump, that is pinned at the origin, remains. In this case, transport is inhibited.
Whenever the initial root density of the two halves is known, GHD provides an exact solution of the partitioning protocol, see Fig. \ref{fig_transport} and SM \cite{Note1} for further evidence.

\begin{figure}[t]
	\includegraphics[width=\linewidth]{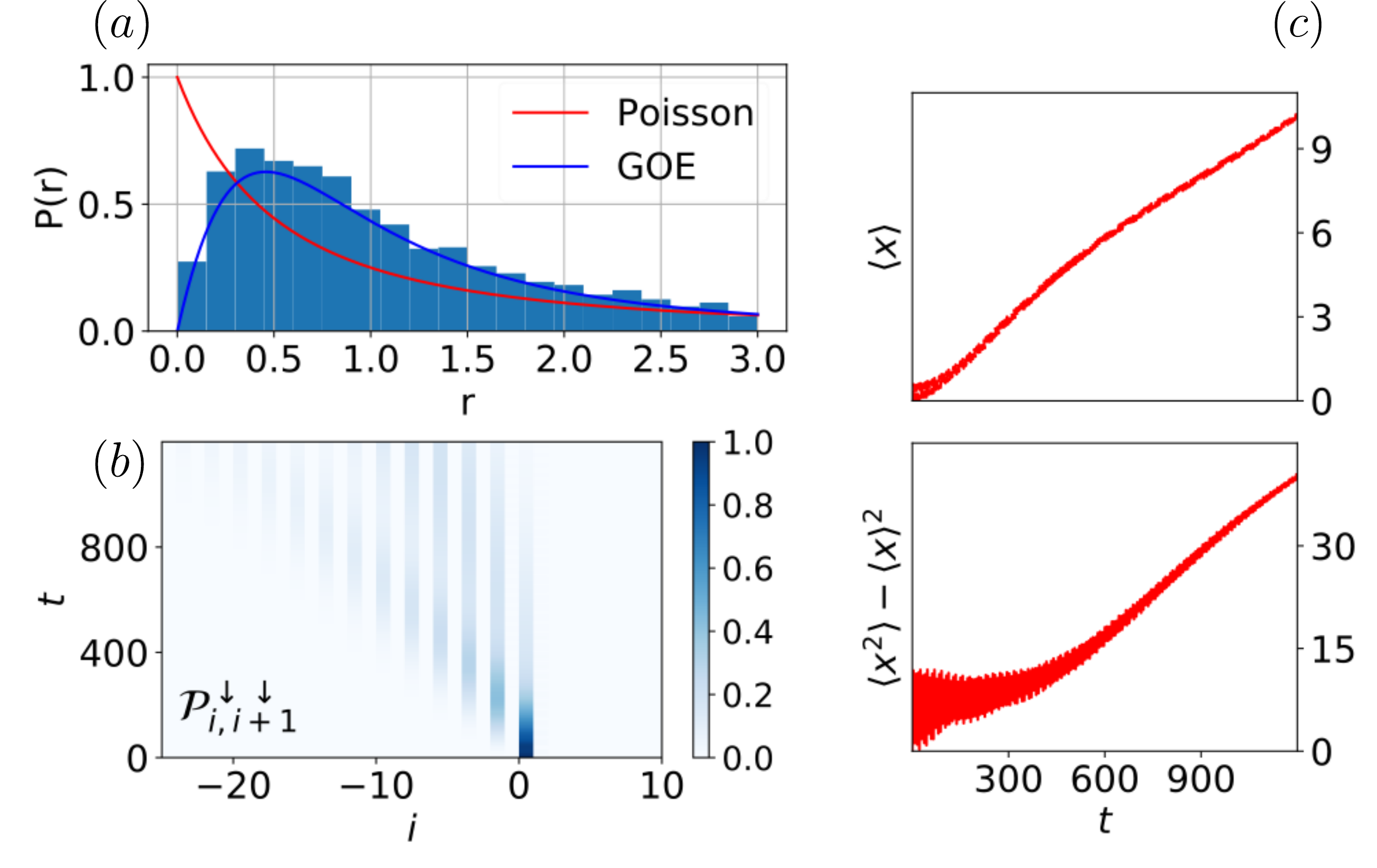}
	\caption{(\textit{a}) The level statistics analysis shows compatibility with the Gaussian Orthongonal Ensamble \cite{PhysRevLett.117.207201}, suggesting that the sector with a two-magnon cluster is not integrable. The distribution function $P(r)$ is defined in \cite{Note1} and computed with the exact diagonalization package QuSpin \cite{Quspin1,Quspin2}. (\textit{b}) A two-magnon cluster, initially at the center of a chain of length $L=80$ bipartitioned into anti-ferromagnetic and ferromagnetic halves, can move to the left by virtue of the magnon-assisted hopping. We track its position by measuring the projector on two consecutive flipped spins $\mathcal{P}_{i,i+1}^{\downarrow\downarrow}$. (\textit{c}) At large times, the position $\langle x\rangle$ and variance $\langle x^2\rangle-\langle x\rangle^2$ of the cluster evolve linearly in time, as described in the main text. The TEBD simulations for (\textit{b}) and (\textit{c}) are done with the Ising Hamiltonian with parameters corresponding to $\Delta = 0.5$ and $\mathcal{J}=-1$.}
	\label{fig_integrabilitybreaking}
\end{figure}

\paragraph{Beyond isolated magnons.---}
Sectors which contain frozen clusters of magnons appear to be generically not integrable: their energy level statistics \cite{PhysRevB.75.155111, PhysRevLett.110.084101} falls into the class of the Gaussian orthogonal random matrix ensemble, see Fig. \ref{fig_integrabilitybreaking} (\textit{a}) and \cite{Note1} for a detailed analysis.
As mentioned before, within leading order perturbation theory clusters are frozen when isolated \footnote{The hopping of an isolated cluster composed of $\ell$ magnons scales $\sim \hpe^{2\ell}$.} and do not contribute to transport by themselves, but their mobility is activated by the scattering with a magnon.
If the scattering is reflective, the cluster stands still, but if transmission occurs the cluster hops by two sites in the direction opposite to the traveling magnon. Therefore, one can relate the cluster displacement $x$ with the total magnetization transported through it as $x=2\delta S^z$. Given that, the cluster position reflects the local transport of spin and its fluctuations.
At late times, a cluster of two magnons undergoes a biased random walk, hopping in the left and right directions with certain rates $R_{L,R}$ which depend on the interactions with the magnonic gas and being proportional to its density. 
Hence, at late time the cluster experiences diffusion \cite{Note1} with a linear growth of the average position and variance, see Fig. \ref{fig_integrabilitybreaking}.

\paragraph{Conclusions and outlook.---}
We discussed the rich phenomenology and transport in the weakly tilted Ising spin chain, exhibiting fragmentation, emergent integrability and magnon-assisted cluster dynamics.
Rydberg atoms in optical tweezers could be used to probe the slow exotic physics of  magnons and clusters discussed here. This experimental platform provides a versatile tool for studying many-body quantum dynamics of Ising-type models in a tilted field \cite{bernien2017probing, PhysRevX.7.041063, PhysRevX.8.021070}. The ability to tune the model parameters and the unprecedented control of the initial state \cite{browaeys2020many} opens a pathway towards experimental investigation of the constrained integrable dynamics emerging in the Ising model in a weak transverse field. In particular, the latter can be seen as a quantum simulator of the Alcaraz Bariev model with completely tunable interaction.
Finally, interesting questions concerning the role of a finite density of integrability-breaking clusters on the late time thermalization and transport remain open for future investigations.

%%%%%%%%%%%%%%%%%%%%%%%%%%%%%%%%%

\paragraph{Acknowledgements.---}We acknowledge useful discussions with Bruno Bertini, Tom Iadecola, Alessio Lerose and Yuan Miao. We thank Bhilahari Jeevanesan for help with the exact diagonalization study of Hilbert space fragmentation.  The work of U.B. and S.M.~is supported by the Emmy Noether Programme of German Research Foundation (DFG) under grant no.~MO 3013/1-1.  
AB acknowledges support from the Deutsche Forschungsgemeinschaft (DFG, German Research Foundation) under Germany's Excellence Strategy–EXC–2111–390814868.
%%%%%%%%%%%%%%%%%%%%%%%%%%%%%%%%%

\bibliography{library}

%%%%%%%%%%%%%%%%%%%%%%%%%%%%%%%%%

\onecolumngrid
%\appendix
\newpage

\setcounter{equation}{0}  
\setcounter{figure}{0}
% reset equation counter
\setcounter{page}{1}
\setcounter{section}{0}    % reset section counter
\renewcommand\thesection{\arabic{section}}    % puts letters as section numbering
\renewcommand\thesubsection{\arabic{subsection}}    % puts letters as section numbering
\renewcommand{\thetable}{S\arabic{table}}
\renewcommand{\theequation}{S\arabic{equation}}
\renewcommand{\thefigure}{S\arabic{figure}}
\setcounter{secnumdepth}{2}  % if the subsections need to be numbered

\begin{center}
{{\large Supplementary Material \\ 
\titleinfo}\\
Alvise Bastianello, Umberto Borla, Sergej Moroz
}
\end{center}

%%%%%%%%%%%%%%%%%%%%%%%%%%%%%%%%%
\section{Second-order effective theory}

Here we derive the effective Hamiltonian by adopting the Schrieffer-Wolff transformation \cite{PhysRev.149.491, bravyi2011schrieffer}. While the unperturbed Hamiltonian $H_0$ trivially preserves the number of magnons $N$ and the number of domain walls $D$, the perturbation $V=- \hpe \sum_i X_i$ changes them. We eliminate transitions that do not conserve $N$ and $D$ order by order in $\hpe$ by performing a unitary transformation of the Hamiltonian
\beq
H_{eff}=e^S H e^{-S}=H+[S, H]+ \frac 1 2 [S,[S, H]]+\dots,
\eeq
where the anti-Hermitian operator $S$ is organized in the power series $S=\sum_{n=1}^\infty S^{(n)}$ in the transverse field coupling $\hpe$.
As a result, the expansion of the effective Hamiltonian in $\hpe$ reads
\beq
H_{eff}=H_0+\underbrace{\Big( [S^{(1)}, H_0]+V\Big)}_{H_{eff}^{(1)}}+ \underbrace{\Big( [S^{(2)}, H_0]+ [S^{(1)}, V] +\frac 1 2 [S^{(1)}, [S^{(1)}, H_0]]  \Big)}_{H_{eff}^{(2)}}+ \dots.
\eeq
Now the terms $S^{(n)}$ are chosen such that up to the $n$-th order in the perturbation coupling $\hpe$ the effective Hamiltonian operates exclusively within the degenerate subspaces of the Hamiltonian $H_0$. Mathematically, one has $[H_{eff}^{(n)}, \mathcal{P}_{N, D}]=0$, i.e. the $n^{th}$ order contribution to effective Hamiltonian $H_{eff}^{(n)}$ commutes with every operator $\mathcal{P}_{N, D}$ that projects on the Hilbert subspace with $N$ magnons and $D$ domain walls.
Since $V$ changes the number of magnons, it is purely off-diagonal. Hence, the linear order Hamiltonian $H_{eff}^{(1)}$ vanishes
\beq
H_{eff}^{(1)}=\mathcal{P}_{N, D} V \mathcal{P}_{N, D}=0.
\eeq
The quadratic order of the effective Hamiltonian is
\beq
H_{eff}^{(2)}= \mathcal{P}_{N,D} \left(\left[S^{(1)}, V\right]+\frac{1}{2}\left[S^{(1)},\left[S^{(1)}, H_{0}\right]\right]\right) \mathcal{P}_{N,D} 
=\mathcal{P}_{N,D} V \frac{1-\mathcal{P}_{N, D}}{E^{(0)}_{N,D}-H_0}  V \mathcal{P}_{N,D}, 
\eeq
where $E^{(0)}_{N,D}$ is the unperturbed energy of the degenerate manifold with $N$ magnons and $D$ domain walls.
 \begin{figure}[b!]
	\includegraphics[width=0.4\linewidth]{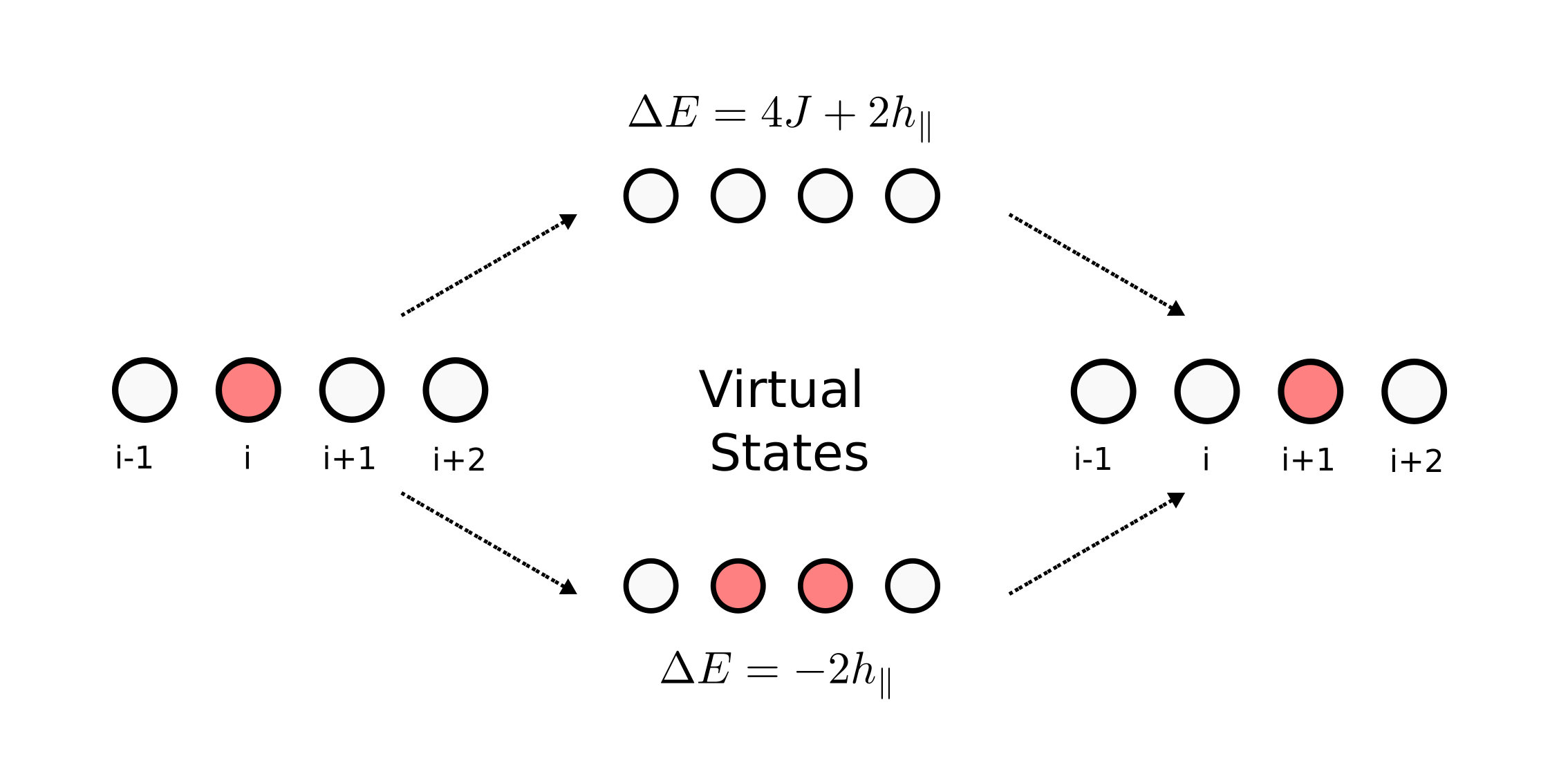}
	\hspace{50pt}
	\includegraphics[width=0.4\linewidth]{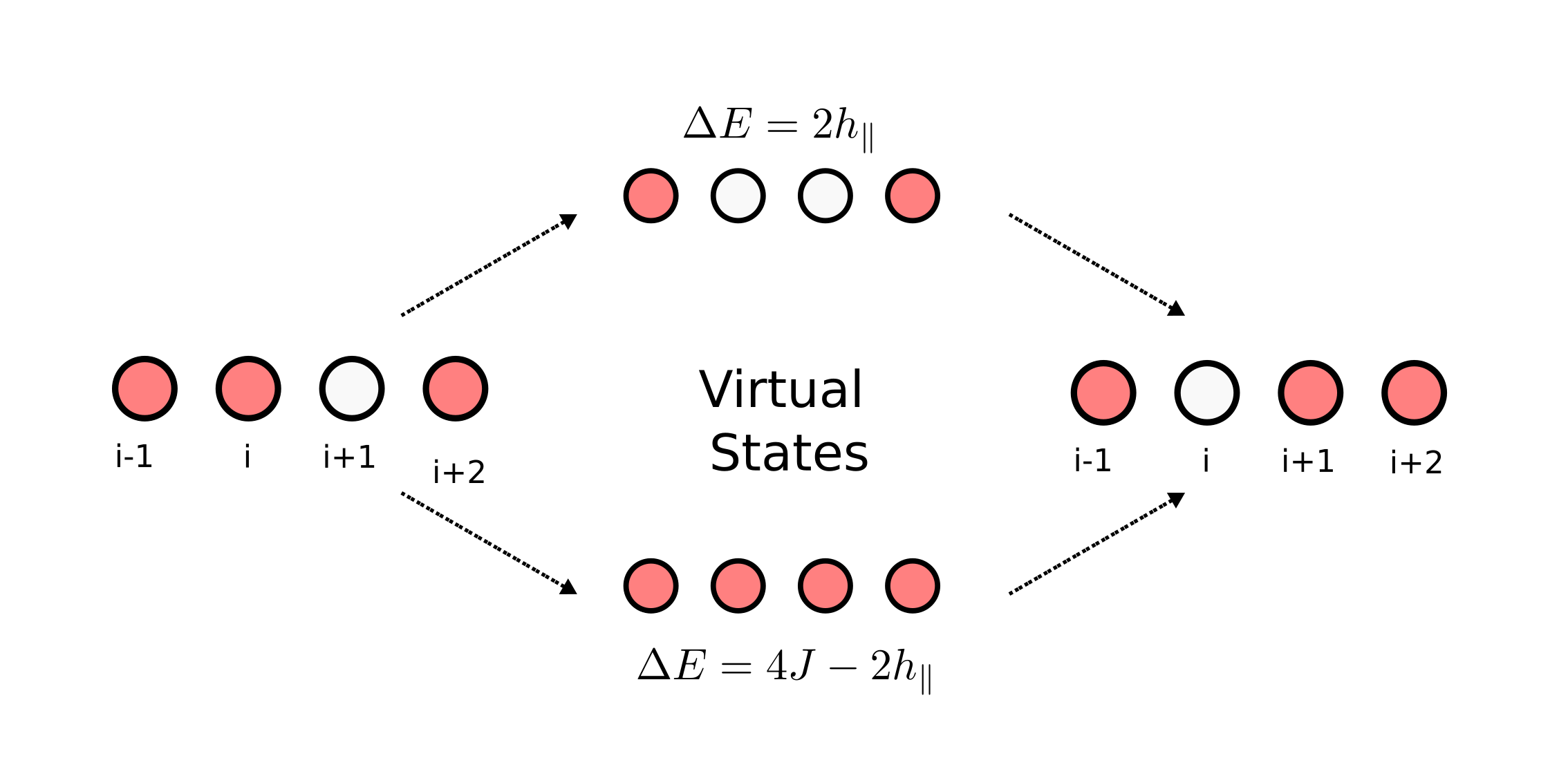}
	\caption{Virtual second-order processes that give rise to the nearest-neighbor spin exchange. }
	\label{figS1}
\end{figure}

As illustrated in Fig. \ref{figS1}, at the second order in $\hpe$, a nearest-neighbor spin-exchange term is generated
\beq \label{SMhop}
\begin{split}
H_{eff, se}^{(2)}
&=-\frac{\hpe^2 J}{\hpa(\hpa+2J)}  \sum_j \mathcal{P}^{+}_{j-1, j+2} \left( S^+_j S^-_{j+1}+h.c.  \right)
-\frac{\hpe^2 J}{\hpa(\hpa-2J)}  \sum_j \mathcal{P}^{-}_{j-1, j+2} \left( S^+_j S^-_{j+1}+h.c.  \right) \\
\end{split}
\eeq 
where
$S^{\pm}_j=(X_j \pm i Y_j)/2$ is the creation/annihilation spin $1/2$ operator on the site $j$. The operator $\mathcal{P}^{\pm}_{i,j}=\left(1\pm (Z_i +Z_j)+ Z_i Z_j \right)/4$
is a projector on spin up-up and down-down pair states, respectively.
Notably, any longer-range spin exchange vanishes because all virtual processes exactly cancel each other in that case.

\begin{figure}[t!]
	\includegraphics[width=0.7\linewidth]{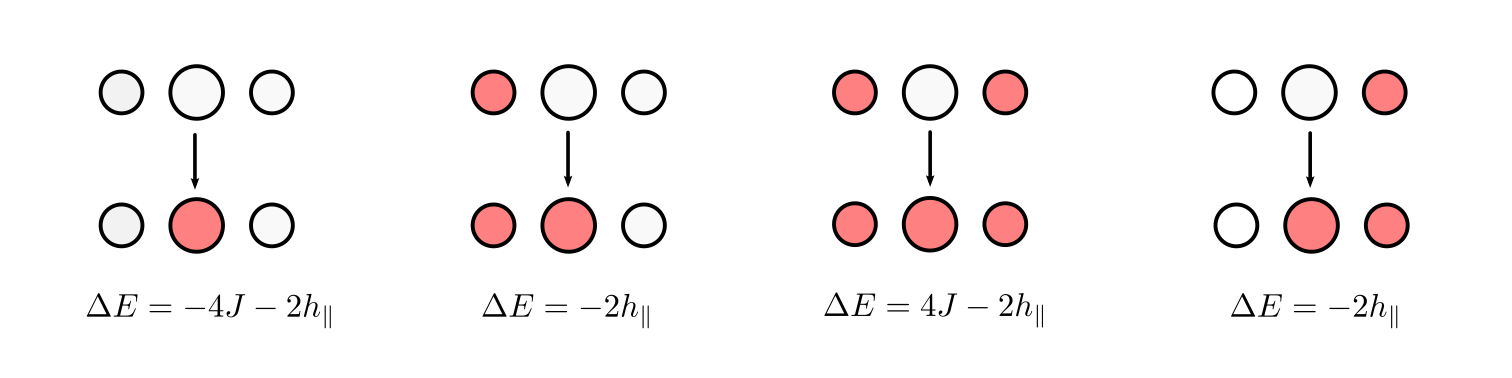}
	\caption{After flipping the middle spin from up to down, a virtual state is obtained whose energy differs by $\Delta E$ from the energy of the original configuration.}
	\label{figS2}
\end{figure}

We derive now the interaction terms generated by the perturbation theory.
To this end, we take into account all second-order processes, where first a spin is flipped by the perturbation $V=- \hpe \sum_i X_i$ and next the very same spin is flipped back again. The energy of the intermediate virtual state depends on the two surrounding spins as illustrated in Fig. \ref{figS2}. We found that all these processes generate the following effective Hamiltonian
\beq \label{SMint}
\begin{split}
H_{eff, int}^{(2)}
&=-\frac{\hpe^2 J^2}{\hpa\alpha}  \sum_j Z_{j-1}Z_{j}Z_{j+1} +\frac{\hpe^2 J}{ \alpha}  \sum_j Z_{j}Z_{j+1} -\frac{\hpe^2 (\hpa^2-2 J^2)}{2\hpa\alpha}  \sum_j Z_{j},
\end{split}
\eeq 
where $\alpha=\hpa^2-4 J^2$. We observe that a three-spin interaction term is generated by the second-order perturbation theory. Moreover, the Ising and longitudinal terms, present in the unperturbed Hamiltonian $H_0$, acquire  small perturbative shifts.

\begin{figure}[t!]
	\includegraphics[width=0.7\linewidth]{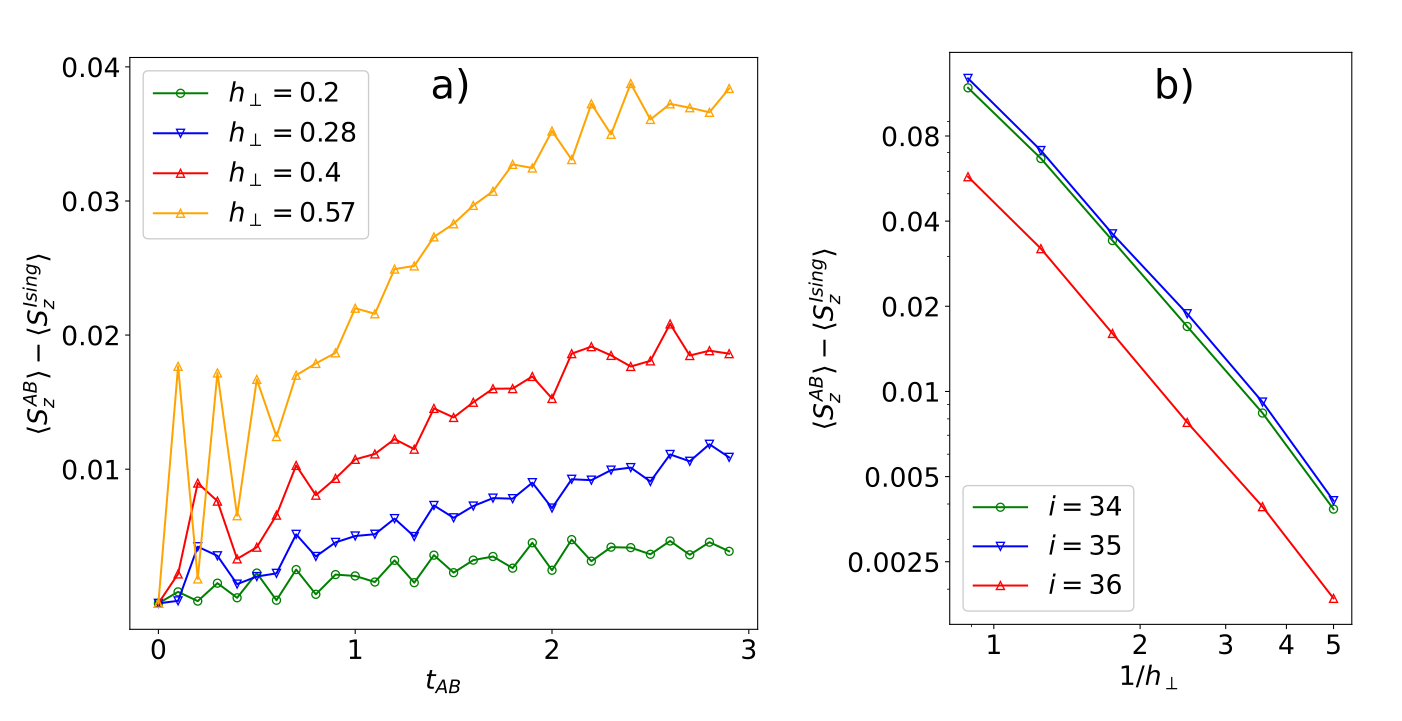}
	\caption{We analyze systematically the discrepancy between the Ising model and the effective Alcaraz-Bariev Hamiltonian derived with second order perturbation theory.
	For this purpose, we consider a partitioning protocol on a chain of $80$ sites, where the first half of the chain $1\le i\le 40$ is initialized in a Neel state, while the rest $40<i\le 80$ is in a fully polarized state.
	a) At site $i=36$  (i.e. four sites left of the junction), the discrepancy in $\langle S_z \rangle$ between the Alcaraz-Bariev and Ising prediction grows approximately linearly on the Alcaraz-Bariev time scale $t_{AB}$. Similar behavior is observed at other sites on the chain. b) At fixed value of the AB time $t_{AB}=3$, we study the convergence in $\hpe$ at different points of the chain in the proximity of the junction. We show that the magnetization $\langle S_z \rangle$ converges as a power-law in $h_{\perp}$ to the value predicted by the effective AB model. A fit of the deviation is compatible with the expected $\propto \hpe^2$ behavior.
	}
	\label{fig_err}
\end{figure}

Putting now the spin-exchange \eqref{SMhop} and the interaction \eqref{SMint} contributions together, we arrive at the complete second-order Hamiltonian \eqref{eff2}. We checked that our final result agrees with the effective Hamiltonian computed in Ref. \cite{PhysRevA.95.023621}.
Before closing this section, we wish to comment on the expected corrections beyond the second order perturbation theory. 
While these additional terms can be explicitly computed by moving to the next order in the Schrieffer-Wolff transformation, the calculation is lengthy and beyond the scope of our work. Hence, we limit ourselves to characterize the scaling.
As already mentioned in the main text, the perturbation induced by the transverse field changes the number of magnons, hence only even orders in the perturbation theory contribute. Therefore, the next-to-leading order correction to the effective Hamiltonian scales as $\mathcal{O}(\hpe^4)$. 
Furthermore, additional contributions emerge due to the fact that the spin degrees of freedom appearing in Eq.\eqref{SMint} are in the Schriffer-Wolff rotated basis. Rotating back to the original spin degrees of freedom, the Pauli matrices get $\propto \hpe^2$ corrections which ultimately result in further $\mathcal{O}(\hpe^4)$ corrections to Eq. \eqref{SMint}.
Given that, at short times, generic observables acquire corrections that grow linearly in time $\mathcal{O}(t \hpe^4)$. However, notice that as $\hpe$ is taken smaller, also the overall energy scale of the effective Alcaraz-Bariev model is reduced as $\hpe^2$. Therefore, for practical purposes one wishes to express the corrections in the limit of small $\hpe$, while keeping constant the timescale in energy-units of the Alcaraz-Bariev model $t_{AB}=\mathcal{J} t\propto t \hpe^2$ with $\mathcal{J}=2\hpe^2 \hpa^{-1} J/(\hpa+2J)$. Following this reasoning, corrections beyond the second order approximation are expected to scale as $\mathcal{O}(t_{AB}\hpe^2)$ at small times, as we indeed observe in Fig. \ref{fig_err}.

%%%%%%%%%%%%%%%%%%%%%%%%%%%%%%%%%

\subsection{Number of sectors in the model \eqref{eff2}}
Here we demonstrate that on a closed chain of a length $L\gg 1$ when expressed in the $Z$-basis the Hamiltonian \eqref{eff2} splits into a large number of independent blocks which grows exponentially with the system size. 

Following ideas from \cite{PhysRevLett.124.207602}, we start from the observation that each independent sector can be labelled by a reference configuration
\beq
\underbrace{\boxed{\text { frozen state }}}_{L-2 k} \underbrace{\boxed{\uparrow \downarrow\uparrow \downarrow \cdots \uparrow \downarrow}}_{2 k}.
\eeq
The frozen state is constructed out of clusters of magnons, but does not contain isolated magnons. The form of the kinetic term in the effective Hamiltonian \eqref{eff2} ensures that in the absence of isolated magnons these clusters are immobile. The number of independent frozen states of a length $l$ follows the Fibonacci recurrence $F_{l+1}=F_l+F_{l-1}$ and thus for $l\gg1$ the number $F_l$ grows exponentially as $\varphi^l$ \cite{PhysRevLett.124.207602}, where $\varphi=(1+\sqrt{5})/2$ is the golden ratio. To estimate the total number $B_L$ of independent blocks of the Hamiltonian \eqref{eff2}, we compute the number of frozen states which can fit into the chain of length $L\gg1$
\beq
B_L=\sum_{k=0}^{L/2}F_{L-2k}\approx \sum_{k=0}^{L/2} \varphi^{L-2k}=\varphi^L \sum_{k=0}^{L/2} \varphi^{2k}\approx \varphi^L \frac{1}{1-\varphi^{-2}}=\varphi^{L+1}.
\eeq

\subsection{The Alcaraz Bariev model as a low temperature limit of the resonant antiferromagnetic Ising chain}
We also note that the Alcaraz-Bariev constraint naturally emerges from the Ising chain at low temperatures at near-resonant couplings. To see that the proximity to the resonance enforces the projector, let us consider the energy of a cluster of $N$ down spins in the sea of up spins: this configuration has (unperturbed) energy $E^{(0)}_{N, 2}=E^{(0)}+2N \hpa +4J$. Imagine now that $\hpa>0$ and one chooses $J=-\hpa/2+\delta $ with $|\hpe|\ll |\delta |\ll |\hpa|$, then clusters of size $N>1$ are thermally  suppressed by the factor $\sim e^{-\beta 2(N-1)\hpa}$. For $\beta\gg\hpa^{-1}$ the thermal state is effectively projected onto the sector of isolated magnons $e^{-\beta H}\to \mathcal{P}_{1}e^{-2 \beta\delta  \sum_j Z_j }\mathcal{P}_{1}$. From the point of view of the AB model, this density matrix describes an infinite-temperature state which can be easily described by integrable techniques.

%%%%%%%%%%%%%%%%%%%%%%%%%%%%%%%%%

\section{The integrable Alcaraz-Bariev model: thermodynamics and hydrodynamics}

In the original paper by Alcaraz and Bariev \cite{Alcaraz_Bariev}, the constrained XXZ model has been solved by coordinate Bethe Ansatz. They studied the following Hamiltonian
\beq 
H_{AB}=- \mathcal{J} \sum_j \mathcal{P}_{T} \left( S^x_j S^x_{j+1}+S^y_j S^y_{j+1}+ \Delta S_j^z S_{j+T}^z   \right) 
\mathcal{P}_{T},
\eeq
With the projector $\mathcal{P}_T$ excluding consecutive magnons (in our notation represented by down spins) that are closer than $T$ sites. The eigenstates of the Hamiltonian (that are also common eigenstates of all conserved quantities) have the explicit representation

\beq\label{eq_CBA}
|\{k_i\}_{i=1}^N\rangle\propto \sum_P\sum_{j_{i+1}-j_i>T} A(P) \exp\left[i\sum_{i=1}^N j_i k_{P(i)}\right]\prod_{i=1}^N \sigma^x_{j_i} |0\rangle
\eeq
where $|0\rangle$ is the reference state with all spins up. Above, the summation is over all possible permutations $P$ of the set of $N$ wavevectors and the constrained summation over the sites $\{j_i\}_{i=1}^N$ is ordered.
The coefficients $A(P)$ encode the effect of interactions and satisfy
\beq
A(P)=-e^{i\Theta(k_{P(j)},k_{P(j+1)})}A(\Pi_{j,j+1} P)
\eeq
with $\Pi_{j,j+1}$ the permutation swapping the elements in position $j$ and $j+1$.
Above, $\Theta(k,q)$ is the scattering phase of the Alcaraz Bariev model in the momentum $k$ space.
It turns out that $\Theta$ is a simple deformation of the scattering phase of the XXZ spin chain \cite{Alcaraz_Bariev}
\beq\label{eq_ThetaAB}
\Theta(k,q)=T(k-q)+\Theta^{XXZ}(k,q)\, ,\hspace{2pc}
\Theta^{XXZ}(k,q)=-i\log\left[  \frac{1-2\Delta e^{ik}+e^{i(k+q)}}{1-2\Delta e^{iq}+e^{i(k+q)}}\right]\, .
\eeq

As we mentioned in the main text, periodic boundary conditions lead to quantization of the momenta encoded in the Bethe equations
\beq\label{eq_BT_s}
e^{i k_j L}= (-1)^{N-1}\prod_{\ell\ne j} e^{i\Theta(k_j,k_\ell)} \, \hspace{2pc}j=\{1,...,N\}\, .
\eeq
Without loss of generality, we assume $N$ to be odd.
The analytic structure of the scattering phase, or better of the scattering matrix $S(k,q)\equiv e^{i\Theta(k,q)}$ is crucial. Indeed, the momenta $k_j$ are not necessarily real, but they can also have a non-trivial imaginary part. 

Let us first consider the large $L$ limit keeping $N$ fixed: for complex solutions, in this limit, $e^{i k_j L}$ is either diverging or vanishing depending on the sign of the imaginary part. On the other side of the equality \eqref{eq_BT_s}, this zero or divergence must be reflected in the scattering matrix. In other words, the imaginary part of the momenta is governed by zeroes or poles in the scattering matrix. Since the factor $e^{iT(k-q)}$ cannot vanish or diverge, the latter are completely dictated by the scattering matrix of the XXZ spin chain $S^{XXZ}(k,q)=\exp[i\Theta^{XXZ}(k,q)]$. 
As a next step, the pattern of the complex solutions obtained in the limit $L\to \infty$ at fixed $N$ is used to build the true thermodynamic limit: this procedure is known as string hypothesis \cite{takahashi2005thermodynamics} and must be taken with a grain of salt. Let us postpone this question and assume its validity. 

It is well-known that the momentum parametrization is not the best to study the analytical properties of the XXZ scattering matrix. A more efficient parametrization $k=p(\lambda)$ is in terms of rapidities $\lambda$.
This parametrization is $\Delta-$dependent and we report it at the end of this section for completeness. In terms of the rapidities, the XXZ scattering phase depends only on their differences $\Theta^{XXZ}(p(\lambda),p(\lambda'))\to \Theta^{XXZ}(\lambda-\lambda')$, where we slightly abuse the notation for the sake of simplicity. In this language, the complex solutions of the Bethe equations in the $L\to \infty$ limit at fixed $N$ can be organized in sets of rapidities with the same real part, but shifted along the imaginary direction. These special solutions are called strings. The energy $\epsilon_j(\lambda)$ (and in general any conserved charge) carried by a string of species $j$, is obtained summing over the constituent of the string $\epsilon_j(\lambda)=\sum_a \epsilon(\lambda+i\delta_a^j)$, where from now on $\lambda$  denotes the real-valued rapidity carried by the string. Moreover, the number of spin flips $m_j$ is nothing else than the number of components of the string. The XXZ strings are well known and for details we refer the reader to Ref. \cite{takahashi2005thermodynamics}.

Usually, the rapidities belonging to the same string are grouped together in the Bethe equations, which then become a set of constraints for the real part of the rapidities and are now called Bethe-Takahashi equations. Carrying out this procedure, one defines the scattering phase of the strings $j$ and $j'$ by summing over their constituents
\beq\label{eq_s13}
\Theta_{j,j'}(\lambda,\lambda')=\sum_{a,a'}\Theta(\lambda+i\delta_a^j,\lambda'+i\delta_{a'}^{j'})=T p(\lambda)m_{j'}-T m_j p(\lambda')+\Theta_{j,j'}^{XXZ}(\lambda-\lambda')
\eeq
The scattering phase of the XXZ spin chain, as well as the momentum of the string $p_j(\lambda)$, the magnetization and all the necessary details are reported at the and of this section.
With $\Theta_{j,j'}(\lambda,\lambda')$ one can apply the Thermodynamic Bethe Ansatz and construct the thermodynamics. 

However, here we would like to point out an important fact: while for $|\Delta|<1$ the string hypothesis determines the full thermodynamics of the XXZ spin chain (and thus of the AB model), this is not the case for $|\Delta|\ge 1$.
In the XXZ model, the string hypothesis covers only one magnetization sector \cite{PhysRevB.96.115124}, in our notation only states with $\langle S^z_j\rangle>0$. In the XXZ case, the way out of this problem is to use the spin reflection symmetry $S_j^z\to -S^z_j$ of the XXZ Hamiltonian. Specifically, to cover the sector with a negative magnetization, one picks as a reference state the ferromagnetic state with all spins down and builds the string hypothesis on top of it. These two symmetric copies of the string hypothesis are distinguished by the magnetization sign $\ff=\pm 1$ that selects which sector one wish to describe.

Importantly, in the AB model this strategy is not feasible, since the $S^z$ reflection symmetry is broken. To access both sectors, in the following we build on the observation of the original AB paper \cite{Alcaraz_Bariev} that the constrained XXZ model can be seen as an ordinary XXZ model in a reduced effective volume.
%%%%%%%%%%%%%%%%%%%%%
\subsection{The AB model as the XXZ chain in reduced volume}

Without loss of generality, we assume $N$ being odd and reorganize Eq. \eqref{eq_BT_s} as
\beq
e^{i k_j(L-TN)}= e^{-iTP}\prod_{\ell\ne j} e^{i\Theta^{XXZ}(k_j,k_\ell)} \, ,
\eeq
where we introduced $P=\sum_j k_j$.
The above can be interpreted as the Bethe equations of a XXZ spin chain in a reduced volume $\tilde{L}=L-TN$ and with periodic boundary conditions twisted by the factor $e^{-iTP}$. This trick has already been noticed by Alcaraz and Bariev \cite{Alcaraz_Bariev} who used it  to construct the coordinate Bether ansatz. We will now use this correspondence to address the thermodynamics and hydrodynamics of the AB model.
In the rapidity parametrization, the density the local conserved charges $\hat{Q}$ (except for the magnetization to be discussed later) is
\beq
\tilde{L}^{-1}\langle \hat{Q}\rangle= \sum_j\int \dd \lambda \, q_j(\lambda) \rho_j^{XXZ}(\lambda)
\eeq
with $q_j(\lambda)$ being called the charge eigenvalue. We explicitly rewrite $\tilde{L}=L(1-nT)$ with $n$ being the density of flipped spins.
Hence, we can write
\beq
L^{-1}\langle \hat{Q}\rangle= \sum_j\int \dd \lambda \, q_j(\lambda) (1-nT)\rho_j^{XXZ}(\lambda)=\sum_j\int \dd \lambda \, q_j(\lambda) \rho_j(\lambda),
\eeq
where we identified the rescaled XXZ root density with the root density of the AB model $\rho_j(\lambda)\equiv(1-nT)\rho_j^{XXZ}(\lambda)$. 
In the sectors where the string hypothesis of the AB model is valid, this correspondence naturally emerges comparing the AB and rescaled XXZ thermodynamics.
Now, we will assume its validity also beyond this case.

Let us now consider the magnetization that was ommited above: in the XXZ model at $|\Delta|\ge 1$ one needs to introduce the magnetization sign \cite{PhysRevB.96.115124}
\beq
\tilde{L}^{-1}\langle S^z_j-1\rangle= \frac{1-\ff}{2}+\sum_j\int \dd\lambda \ff|m_j^{XXZ}|\rho_j^{XXZ}(\lambda).
\eeq
Now, we rewrite $\tilde{L}^{-1}\langle S^z_j-1\rangle=\tilde{L}^{-1} Ln=n(1-nT)^{-1}$ and solve the above for $n$
\beq
n=\frac{(1-\mathfrak{f})}{2+T(1-\mathfrak{f})}+\sum_j \int \dd\lambda \frac{2\mathfrak{f}}{2+T(1-\mathfrak{f})}|m_j^{XXZ}| \rho_j(\lambda).
\eeq
This leads to the natural identification $m_j\equiv (1+T(1-\ff)/2)^{-1} m_j^{XXZ}$ that we have already anticipated in the main text.
The correspondence is then easily extended to the whole thermodynamics. In particular, the definition of the total root density
\beq\label{eq_rhot}
\sigma_j\rho^t_j(\lambda)=\frac{2}{2+T(1-\ff)}\frac{\partial_\lambda p_j(\lambda)}{2\pi}-\sum_{j'}\int \frac{\dd\lambda}{2\pi}\partial_\lambda \Theta_{j,j'}(\lambda,\lambda')\rho_{j'}(\lambda)\, ,
\eeq
where $\Theta_{j,j'}$ is defined in Eq. \eqref{eq_s13} is consistent with the expected rescaling $\rho^t_j(\lambda)\equiv(1-nT)[\rho_j^t]^{XXZ}(\lambda)$.

Finally, let us address the problem of constructing thermodynamics of thermal states in the presence of a magnetic field $e^{-\beta(H+B \sum_j S^z_j)}$, where $\beta$ denotes the inverse temperature. By means of standard TBA techniques, the root densities of thermal states can be found solving the following integral equation
\beq\label{eq_th_s}
\varepsilon_j(\lambda)=\beta (\epsilon_j(\lambda)-B m_j)-\sum_{j'}\int \frac{\dd \lambda'}{2\pi}\partial_{\lambda'}\Theta_{j,j'}(\lambda,\lambda')\sigma_{j'}\log(1+e^{-\varepsilon_{j'}(\lambda')})\, .
\eeq
with $\rho_j(\lambda)=\rho_j^t(\lambda)(1+e^{\varepsilon_j(\lambda)})^{-1}$ and $\epsilon_j(\lambda)$ the energy of the string, which is the same as the XXZ spin chain. These TBA equations are consistent with first solving TBA equations in the XXZ spin chain in a reduced volume and then taking the proper rescaling afterwards.
Notice that the ferromagnetic spin up state and Neel state are nothing else than ground states ($\beta\to \infty)$ of the AB Hamiltonian with $B=-\infty$ and $B=+\infty$ respectively. Therefore, these states can be easily described with the above equation.
The ferromagnetic spin up state is nothing else than the vacuum hence $\rho_j(\lambda)=0$ (and $\ff=1$ for $|\Delta|>1$), the description of the Neel state depends on $\Delta$.
Indeed, if $|\Delta|<1$ the associated root density is non-trivial, but whenever $|\Delta|>1$ one gets again $\rho_j(\lambda)=0$, but $\ff=-1$.

%%%%%%%%%%%%%%%%%%%%%%%%%%%%%%
\subsection{The hydrodynamics: two equivalent formulations}
In the case of homogeneous interactions, but inhomogeneous state, the hydrodynamics of integrable models is described by the continuity equation \cite{PhysRevX.6.041065,PhysRevLett.117.207201}
\beq\label{eq_ghd_rho}
\partial_t \rho_j(\lambda)+\partial_x(v_j^\text{eff} (\lambda)\rho_j(\lambda))=0
\eeq
or, equivalently,
\beq\label{eq_ghd_fill}
\partial_t \vartheta_j(\lambda)+v_j^\text{eff} (\lambda)\partial_x\vartheta_j(\lambda)=0
\eeq
with $\vartheta_j=\rho_j/\rho^t_j$ being called the filling fraction and the effective velocity defined in the main text, see Eq. \eqref{eq_veff}. It is worth emphasizing that the intuitive expression for the effective velocity $v^\text{eff}_j(\lambda)=(\partial_\lambda\epsilon_j)^\text{dr}/(\partial_\lambda p_j)^\text{dr}$ is not equivalent to Eq. \eqref{eq_veff} in the AB model.
The equivalence between the two equations (\ref{eq_ghd_fill},\ref{eq_ghd_rho}) is not trivial and it requires some formal manipulations to be presented here.

Following Refs. \cite{PhysRevX.6.041065,PhysRevLett.117.207201}, let us demonstrate explicitly how Eq. \eqref{eq_ghd_fill} follows from \eqref{eq_ghd_rho}.
First, one rewrites Eq. \eqref{eq_ghd_rho} as $\partial_t [\vartheta_j(\lambda)\rho_j^t(\lambda)]+\partial_x [\frac{\sigma_j}{2\pi}(\partial_\lambda\epsilon_j)^\text{eff}\vartheta_j(\lambda)]=0$ and expands the derivatives 
\beq\label{eq_39}
\rho_j^t(\lambda)\left[\partial_t \vartheta_j(\lambda)+v^\text{eff}_j(\lambda)\partial_x \vartheta_j(\lambda)\right]+\frac{\sigma_j}{2\pi}\vartheta_j(\lambda)\left[\partial_t (2\pi\sigma_j\rho_j^t(\lambda))+\partial_x (\partial_\lambda\epsilon_j)^\text{eff}\right]=0\, .
\eeq
Next, the second term of the previous equation will be now shown to vanish. First, we take the time derivative of Eq. \eqref{eq_rhot}
\beq\label{eq_40}
\partial_t[2\pi\sigma_j\rho_j^t(\lambda)]=\partial_t\left[-\frac{T(1-\mathfrak{f})}{2+T(1-\mathfrak{f})}\right]\partial_\lambda p_j(\lambda)-\sum_{j'}\int \dd\lambda'\, \partial_t[\partial_{\lambda}\Theta_{j,j'}(\lambda,\lambda')]\rho_{j'}(\lambda')-\sum_{j'}\int \dd\lambda'\, \partial_{\lambda}\Theta_{j,j'}(\lambda,\lambda')\partial_t\rho_{j'}(\lambda').
\eeq
In this model, the presence of the magnetization sign in the string scattering phase gives a time and space depencence to the latter $\partial_t \partial_{\lambda'}\Theta_{j,j'}(\lambda,\lambda')=\partial_\lambda p_j(\lambda)T\partial_t m_{j'} $ and  $\partial_x \partial_{\lambda'}\Theta_{j,j'}(\lambda,\lambda')=\partial_\lambda p_j(\lambda)T\partial_x m_{j'} $. Using these identities and plugging the hydrodynamic equation $\partial_t \rho_j(\lambda)=-\partial_x (v^\text{eff}_j\rho_j)$ in the last term of Eq. \eqref{eq_40} one finds
\beq
\partial_t[2\pi\sigma_j\rho_j^t(\lambda)]=-[\partial_t n+\partial_x j_n]T\partial_\lambda p_j(\lambda)+\\
\partial_x\left[\sum_{j'}\int \frac{\dd\lambda'}{2\pi}\, \partial_{\lambda}\Theta_{j,j'}(\lambda,\lambda')\sigma_{j'}(\partial_\lambda\epsilon_{j'})^\text{dr} \vartheta_{j'}(\lambda)\right]\, .
\eeq
The spin flip continuity equation causes the first term to vanish, while in the second term one recognizes the definition of the dressed derivative of the energy
\beq
\partial_t[2\pi\sigma_j\rho_j^t(\lambda)]=
\partial_x\left[\partial_\lambda \epsilon_j-(\partial_\lambda \epsilon_j)^\text{dr}\right]=-\partial_x(\partial_\lambda \epsilon_j)^\text{dr}\, .
\eeq
This result ensures that the second term in Eq. \eqref{eq_39} indeed vanishes and thus we end up with Eq. \eqref{eq_ghd_fill}.

\subsection{The partitioning protocol}
\label{sec_part_prot}

\begin{figure}[t!]
	\includegraphics[width=0.6\linewidth]{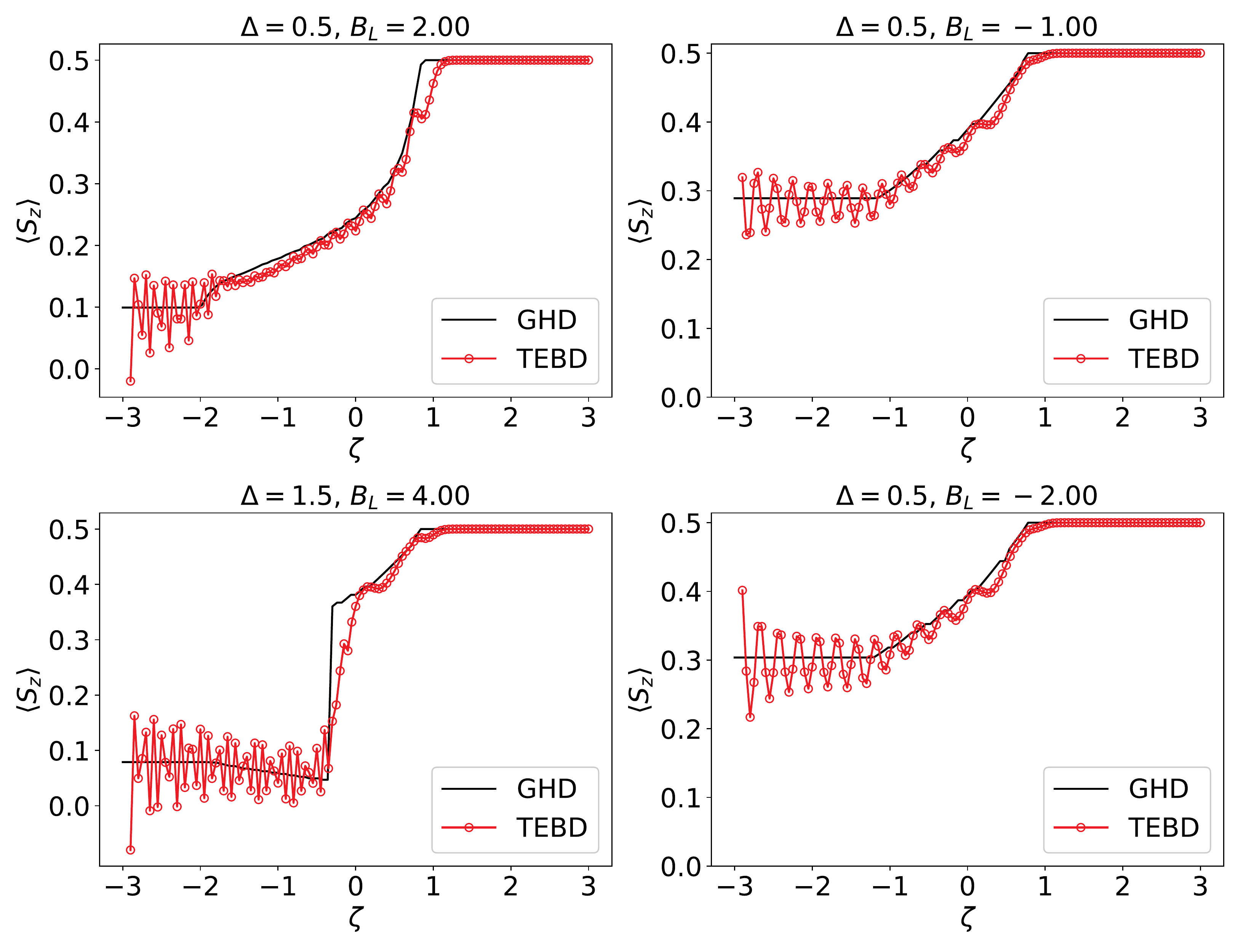}
	\caption{We compare the numerical simulations of partitioning protocols in the Alcaraz Bariev model against the exact hydrodynamic solution. We initialize the state in two halves $|\text{GS}_{\langle Z\rangle}\rangle\otimes|\text{ferro}\rangle$ with $|\text{GS}_{\langle Z\rangle}$ the ground state at fixed magnetization.
	In practice, $|\text{GS}_{\langle Z\rangle}\rangle$ can be obtained by applying an external magnetic field in the $z$ direction ($B_L$ in the figure), as described in Eq. \eqref{eq_th_s}. The same method is used also to obtain the desired matrix product state state in the microscopic simulations. Notice the sharp jump for $\Delta=1.5$, since states above and below half filling are connected. In all the plots we set $\mathcal{J}=-1$.
	}
	\label{fig_ab_ghd}
\end{figure}
The partitioning protocol is best addressed by means of the hydrodynamic equations in the form \eqref{eq_ghd_fill}. In this protocol, the two halves are initializes in two homogeneous states
\beq
\vartheta_j(\lambda)\Big|_{t=0,x}=\theta(x) [\vartheta_j(\lambda)]_R+\theta(-x) [\vartheta_j(\lambda)]_L
\eeq
with $\theta(x)$ the Heaviside theta function $\theta(x>0)=1$ and zero otherwise. The left and right fillings $[\vartheta_j(\lambda)]_{L,R}$ are the initial conditions and must be given as an input. In our case, we probed filling fractions belonging to the class of thermal states defined through Eq. \eqref{eq_th_s}.
Due to the appearance of only first derivatives in the hydrodynamic equation, signaling the ballistic transport, the solution of Eq. \eqref{eq_ghd_fill} with these initial condition is scale-invariant. Namely, for $t>0$ the filling is not an independent function of time and space, but a function of their ratio. We define the ray $\zeta=x/t$ and Eq. \eqref{eq_ghd_fill} admits the solution \cite{PhysRevX.6.041065,PhysRevLett.117.207201}
\beq\label{eq_sol_part}
\vartheta_j(\lambda)=\theta(\zeta-v^\text{eff}_j(\lambda)) [\vartheta_j(\lambda)]_R+\theta(v^\text{eff}_j(\lambda)-\zeta) [\vartheta_j(\lambda)]_L\, .
\eeq

where the $\zeta-$dependence of $v^\text{eff}_j(\lambda)$ is left implicit. Since $v^\text{eff}$ depends on the state through the dressing, the above solution is only implicit and cannot be further analytically simplified. However, very simple recursive numerical schemes guarantee fast convergence: first, one finds an initial ansatz for $\vartheta_j(\lambda)$ ignoring the dressing in the effective velocities in Eq. \eqref{eq_sol_part}. Then, the filling fraction is used to recalculate $v^\text{eff}$ and the procedure is iterated until convergence is reached, which usually happens after only few steps.
In the case where the two halves are initialized in opposite magnetic sectors, one must supplement Eq. \eqref{eq_sol_part} with the proper equation for the sign $\ff$, similarly to what has been done in XXZ \cite{PhysRevB.96.115124}. Imposing spin conservation  $\partial_t n+\partial_x j_n=0$ in the scaling form, one readily obtains an equation similar to \eqref{eq_sol_part}
\beq\label{eq_part_sign}
\ff= \theta(\zeta-\bar{v})\ff_R+\theta(\bar{v}-\zeta) \ff_L
\eeq
with $\ff_{R,L}$ set by the initial conditions and
\beq\label{eq_barv}
\bar{v}=\left(\sum_j \int \dd\lambda  m_j  v_j^\text{eff}(\lambda)\rho_j(\lambda)\right)\left(\frac{1}{2+T}-\sum_j \int \dd\lambda  m_j \rho_j(\lambda)\right)^{-1}\, .
\eeq

\begin{figure}[t!]
	\includegraphics[width=0.7\linewidth]{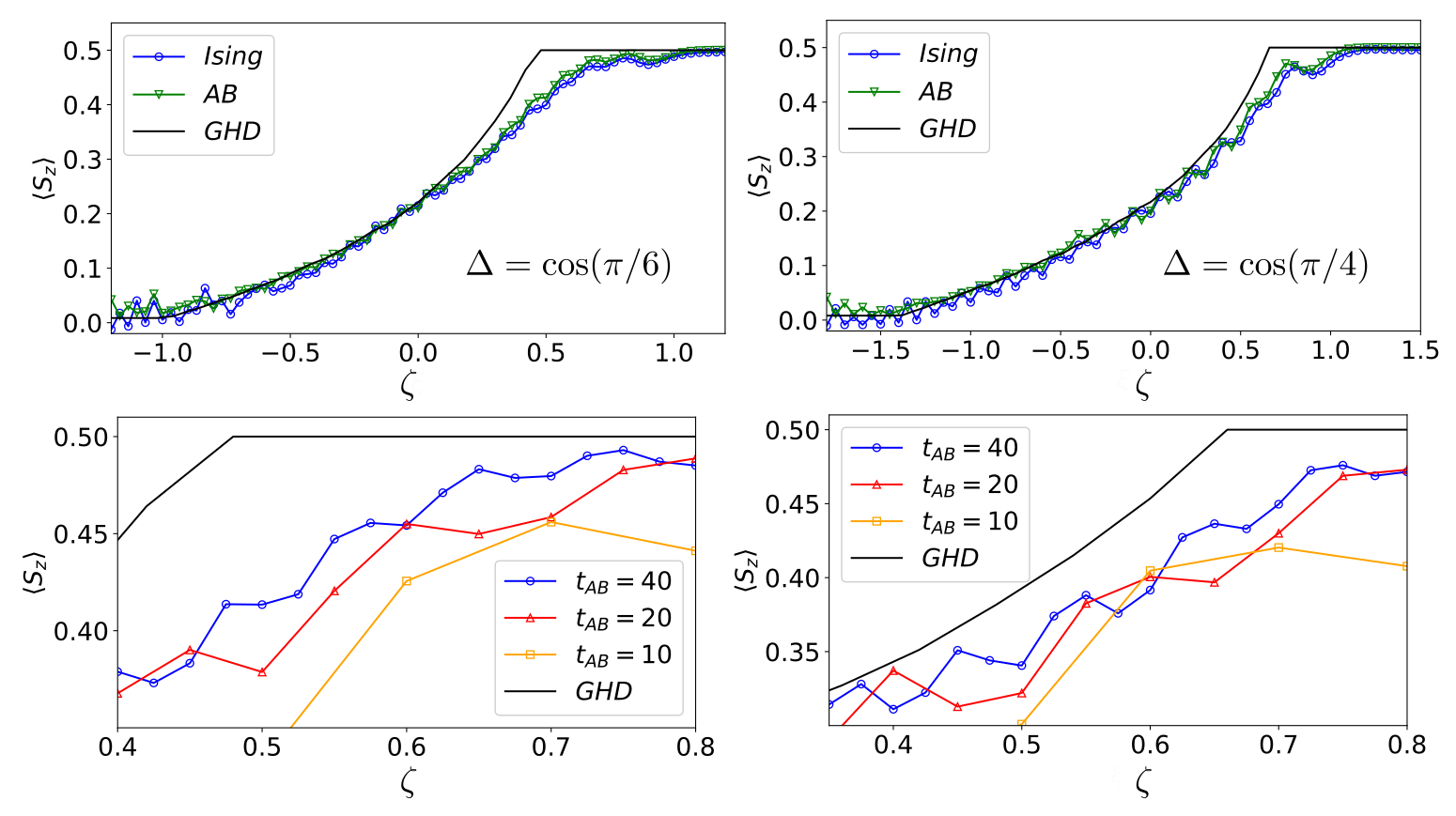}
	\caption{Here we provide further evidence that the Alcaraz-Bariev model gives an effective description of the Ising model in the limit of weak transverse field. Due to its integrability, at large times we can compare expectation values of observables to the ones predicted by the Generalized Hydrodynamics. 
	We consider partitioning protocols where the left half is initialized in the Neel state, while the right half is a fully polarized state. We consider two different values of the AB interaction parameter $\Delta$.
	In the upper panels, we show that the expectation values $\langle S_z \rangle$ at a large time $t_{AB}=30$ in the Ising and AB models match with very good agreement for $\hpe=0.5$, and approach the GHD curve.
	A deviation from the hydrodynamic prediction is evident, but this is a finite-time effect.
	In the lower panels, we zoom on the region where the mismatch is more evident and show that for large times the curves slowly converge to the GHD (only the AB prediction is shown).
	}
	\label{fig_compare}
\end{figure}

Two crucial observations must be made. First, in contrast with the XXZ model, Eq. \eqref{eq_sol_part} depends on the magnetization sign $\ff$ through the scattering phase. Hence,  Eq. \eqref{eq_sol_part} and Eq. \eqref{eq_part_sign} must be solved simultaneously.
Because this dependence, it is not a priory obvious why the value of $\bar{v}$ should not change if one computes it using in Eq. \eqref{eq_barv} the root densities for $\zeta=\bar{v}+0^+$ or $\zeta=\bar{v}+0^-$: the convergence of the iterative solution is rooted on this fact. Indeed, besides the convergence, we also checked the equivalence of the two limits a posteriori: this is a highly non trivial check of the consistency of our solution.
In Fig. \ref{fig_ab_ghd} we provide further checks of the hydrodynamic solution against the TEBD numerical simulation of the Alcaraz-Bariev model, finding excellent agreement as expected.
In Fig. \ref{fig_compare}, we supplement the plots shown in the main text with a further comparison between the Ising dynamics, the Alcaraz-Bariev model and the GHD of the latter.

%%%%%%%%%%%%%%%%%%%%%%%%%%%%%%%%%%%%%%%%%

\subsection{Summary of XXZ thermodynamics}

For completeness, we provide a short summary of the XXZ thermodynamics on which the solution of the AB model is built. For a more extended discussion, we refer to Ref. \cite{takahashi2005thermodynamics}. The sectors with opposite interaction signs are unitary equivalent, hence as customary we focus on the regime $\Delta>0$.

\begin{itemize}
    \item  The case $\Delta\ge 1$:
    The interaction is conveniently parametrized as $\Delta=\cosh\theta$, the string parametrization and scattering phases are
    \beq
p(\lambda)=-i \log\left[\frac{\sin(\lambda-i\theta/2)}{\sin(\lambda+i\theta/2)}\right]\, , \hspace{2pc} \Theta^{XXZ}(\lambda)=-i\log\left[-\frac{\sin(\lambda+i\theta)}{\sin(\lambda-i\theta)}\right]\, .
\eeq
In this sector, the system has infinitely many strings of species $j=\{1,2,...\}$ and the rapidities of the constituents of a string with real rapidity $\lambda$ are obtained by shifting in the imaginary direction
\beq
\lambda^{a,j}=\lambda+i\theta \frac{(j-1-2a)}{2}\, ,\hspace{1pc} a=\{0,...,j-1\}.
\eeq
The scattering phase is given by Eq. \eqref{eq_s13}. In particular, one finds
\beq
\partial_\lambda \Theta_{j,j'}^{XXZ}(\lambda)=(1-\delta_{j,j'})f_{|j-j'|}(\lambda)+f_{j+j'}(\lambda)+2\sum_{s=1}^{\min(j,j')-1}f_{|j-j'|+2s}(\lambda)
\eeq
with
\beq
f_j(\lambda)=\frac{1}{2\pi}\partial_\lambda p_j(\lambda)=\frac{1}{\pi}\frac{\sinh(j\theta)}{\cosh(j\theta)-\cos(2\lambda)}
\eeq
and $\epsilon_j(\lambda)=\mathcal{J}\pi \sinh\theta f_j(\lambda)$ and $|m_j^{XXZ}|=j$. In this sector, the parity of the string is always positive $\sigma_j=1$ and the rapidities of the strings live within a finite domain $\lambda\in [-\pi/2,\pi/2]$. The choice of the magnetization sector $\ff=\pm 1$ only changes the sign of $m_j^{XXZ}$ and nothing else.
\item The case $0<\Delta<1$: 
With the parametrization $\Delta=\cos(\pi\gamma)$ one has
\beq
p(\lambda)=-i\log\left[\frac{\sinh(\lambda+i\pi\gamma/2)}{\sinh(\lambda-i\pi\gamma/2)}\right]\, ,\hspace{2pc}\Theta^{XXZ}(\lambda)=-i\log\left[\frac{\sinh(\lambda-i\pi\gamma/2)}{\sinh(\lambda+i\pi\gamma/2)}\right]\, 
\eeq

The string content depends on the continued fraction representation of $\gamma$
\beq\label{eq_cf}
\gamma=\frac{1}{n_1+\frac{1}{n_2+...}}
\eeq
where $n_i$ are suitable positive integers
and the total number of strings is $\sum_i n_i$.
The constituents of a string of species $j$ carry rapidities
\beq
\lambda^{a,j}=\lambda+i\frac{\pi\gamma}{2}(m_j+1-2a)+i\pi(1-v_j)/4\, , \hspace{2pc} a=\{1,..,m_j\},
\eeq
where the real rapidity $\lambda$ covers the entrire real axis $\lambda\in(-\infty,\infty)$.
The value of the magnetization $m_j$, the parity $\sigma_j$ and the parameter $v_j$ depend on the continued fraction expansion \eqref{eq_cf}. In the simplest case where one chooses $\gamma=1/\ell$, one has $\ell$ strings and
\beq
m_j=j\,,\hspace{1pc} \sigma_j=1\,,\hspace{1pc} v_j=1\,,\hspace{1pc} j<\ell
\hspace{2pc}\text{and}\hspace{1pc}
m_\ell=1\,,\hspace{1pc} \sigma_\ell=-1\,,\hspace{1pc} v_\ell=-1.
\eeq

For the general case, we refer to Ref. \cite{takahashi2005thermodynamics}. Finally, the string scattering data are
\beq
\partial_\lambda\Theta^{XXZ}_{j,j'}(\lambda)=(1-\delta_{m_j,m_{j'}}) a^{v_j v_{j'}}_{|m_j-m_{j'}|}(\lambda)+a^{v_j v_{j'}}_{m_j+m_{j'}}(\lambda)+2\sum_{s=1}^{\min(m_j,m_{j'})-2} a^{v_j v_{j'}}_{|m_j-m_{j'}|+2s}(\lambda),
\eeq
where
\beq
a_x^y(\lambda)=\frac{y}{\pi}\frac{\sin(\pi\gamma x)}{\cos(2\lambda)-y\cos(\pi \gamma x)}\,, \hspace{1pc}
\frac{1}{2\pi}\partial_\lambda p_j(\lambda)= a_{m_j}^{v_j}(\lambda) \,, \hspace{1pc} \epsilon_j(\lambda)=\mathcal{J}\pi \sinh(\pi\gamma) a_{m_j}^{v_j}(\lambda).
\eeq

\end{itemize}

%%%%%%%%%%%%%%%%%%%%%%%%%%%%%%%%%%%%%
\section{The level spacing statistics analysis}
Here we investigate numerically the level spacing statistics of the second-order perturbative effective Hamiltonian \eqref{eff2} that captures the physics of the tilted Ising chain in the regime of a weak transverse field. As argued in the main text, this model exhibits fragmentation of the Hilbert space as the latter splits in the local $Z$-basis into $\varphi^{L+1}$ independent blocks for $L\gg 1$. Given that, we investigate the level statistics of several  large sectors and use it as a diagnostics of integrability of the corresponding sectors.  

\begin{figure}[]
	\includegraphics[width=0.4\textwidth]{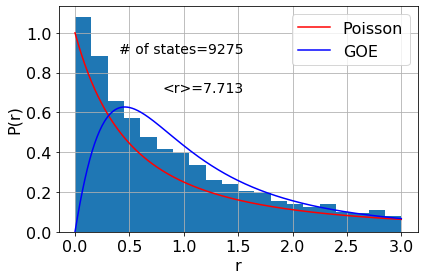}
	\caption{The distribution $P(r)$ of the ratios $r$ of consecutive level spacings of the second-order effective Hamiltonian \eqref{eff2} in the sector with $N_m=5$ isolated magnons on a closed chain of length $L=40$ in the momentum sector with $k=7\times 2\pi/L$. The Ising parameters that fix all parameters of the effective model are $J=1.17$, $\hpa=0.91$ and $\hpe=0.0291$. }
	\label{fig_qs_Poisson}
\end{figure}

We start with a closed chain of length $L=40$ and compute numerically using QuSpin python package \cite{Quspin1, Quspin2} the energy spectrum of the sector populated with $N_m=5$ isolated magnons with momentum $k=7\times 2\pi/L$. Instead of looking directly at the energy level spacings, we follow ideas from \cite{PhysRevB.75.155111, PhysRevLett.110.084101} and compute the ratios of consecutive level spacings $r_n=(E_{n+1}-E_n)/(E_n-E_{n-1})$. The resulting distribution $P(r)$, plotted in Fig. \ref{fig_qs_Poisson}, agrees with $P(r)=1/(1+r)^2$ \cite{PhysRevLett.110.084101} which one gets if the energy levels are completely random  (the Poisson distribution). As a result, our numerics is consistent with integrability of the sectors with only isolated magnons that we argued for in the main text.

\begin{figure}[]
	\includegraphics[width=0.7\textwidth]{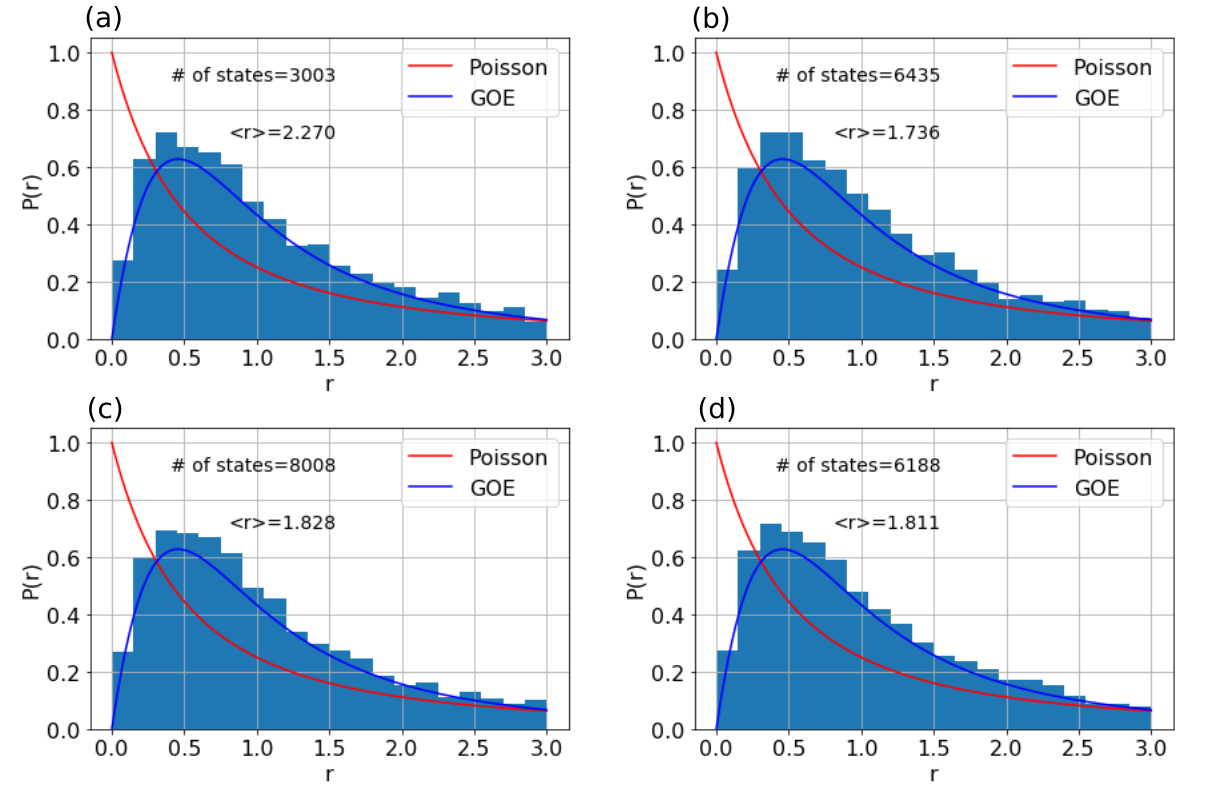}
	\caption{The distribution $P(r)$ of the ratios $r$ of consecutive level spacings of the second-order effective Hamiltonian \eqref{eff2} in the sector with $N_m=10$ magnons including one (a) dimer, (b) trimer, (c) tetramer and (d) pentamer on a closed chain of length $L=25$ in the momentum sector with $k=7\times 2\pi/L$. The Ising parameters that fix all parameters of the effective model are $J=1.17$, $\hpa=0.91$ and $\hpe=0.0291$.}
	\label{fig_qs_GOE}
\end{figure}

We turn now to sectors with clusters. In particular, we consider a closed chain of length $L=25$ with $N_{m}=10$ magnons among which there is one cluster of sizes two, three, four and five, respectively. The resulting distributions of $P(r)$ are plotted in Fig. \ref{fig_qs_GOE}. Since they all plummet at low $r$, the energy levels repel implying that these sectors are not integrable. In fact, all averages $\langle r \rangle$ are not far from the value $1.75$ which is expected for the Gaussian orthogonal ensemble \cite{PhysRevLett.110.084101}.

%%%%%%%%%%%%%%%%%%%%%%%%%%%%%%%%%%%%

\section{Late time dynamics of the smallest clusters}

As we discussed in the main text, the presence of clusters composed  of neighboring magnons breaks integrability and the analytical methods of generalized hydrodynamics. Here we develop a simple phenomenological description to capture the late time dynamics of the clusters.

\begin{figure}[h]
	\includegraphics[width=0.3\textwidth]{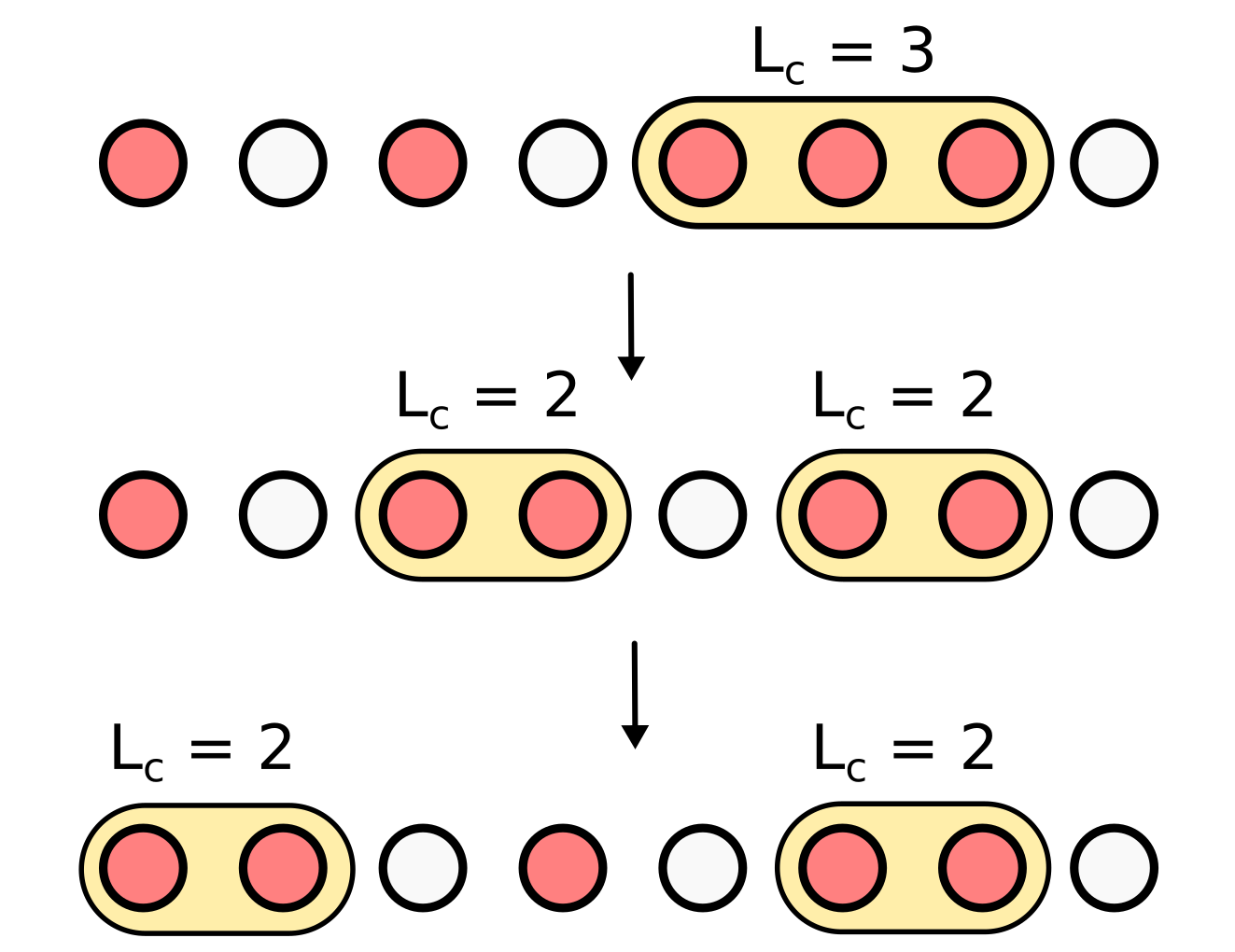}
	\caption{Decay of a three-magnon cluster in two clusters mediated by the interaction with surrounding isolated magnons. }
	\label{fig_assisted_processes}
\end{figure}

Motivated by Fig. \ref{fig_integrabilitybreaking}, let us consider an initial inhomogeneous state in the form of a partitioning protocol $|\Psi_L\rangle \otimes |\Psi_r\rangle$ and, in addition, we place a cluster composed of two magnons at the origin. Clusters in isolation are static in perturbation theory, but the surronding isolated magnons can activate their dynamics. A two-magnon cluster undergoes assisted hopping of two sites at once, mediated by the scattering with an isolated magnon.
The case of a bigger cluster of length $L_c>2$ is more complicated, since they can also decay into smaller clusters at intermediate stages, see Fig. \ref{fig_assisted_processes}.
For the sake of simplicity, we focus here on the case $L_c=2$ that cannot decay into smaller clusters.

For $t>0$, we investigate activation of transport on the timescale where the effective perturbative Hamiltonian is valid.
Far from the cluster, the dynamics is locally integrable and can be rightfully assumed to be described by the GHD equation $\partial_t \rho_j+\partial_x (v_j^\text{eff}\rho_j)=0$. In this perspective, the cluster plays the role of a \emph{dynamical impurity} for the integrable excitations and sets the proper boundary conditions in the form of a generalized scattering matrix. Finding the exact boundary conditions is a challenging problem, since one needs to solve the non-integrable magnon-cluster scattering. Nevertheless, after a transient time, the cluster will be surrounded by a state that reached a local (generalized) equilibrium, hence the interactions between the cluster and the surrounding magnons will remain constant in time. Let us consider the motion of the cluster in a semiclassical approximation, by denoting with $P_t(j)$ the probability of finding the cluster at a position $j$. At any time, the cluster can jump to the left by two sites with rate a $R_L$ and to the right with a rate $R_R$. These rates originate from the interaction with the surrounding isolated magnons: their computation is a fomidable task, but in the present calculation we will treat them as phenomenological parameters constant in time.
Given that, one expects $P_t(j)$  to obey the difference equation
\beq\label{eq_micro_P}
\partial_t P_t(j)=R_L P_t(j+2)+R_R P_t(j-2)-(R_L+R_R)P_t(j).
\eeq
This equation can be easily solved by passing to the Fourier space
\beq
P_t(j)=\sum_{j'}G_{j-j'}(t) P_0(j')\, \hspace{2pc} G_{j}=\int \frac{\dd k}{2\pi}e^{ikj-tR_L(1-e^{i2k})-tR_R(1-e^{-i2k})}\, .
\eeq
At late times when $P_t(j)$ becomes a smooth function of $j$, we can replace discrete jumps with spatial derivatives. As a result, a a biased diffusive equation is obtained
\beq
\partial_t P_t(x)\simeq
2(R_L-R_R)\partial_j P_{t}(j)+2 (R_L+R_R)\partial_j^2 P_t(j)+\mathcal{O}(\partial_j^3 P)\, .
\eeq
From this equation we find for the average displacement $\langle x\rangle=2t (R_R-R_L)$ and its variance $\langle (x)^2\rangle-\langle x\rangle^2=4t (R_R+R_L)$. Remarkably, the expressions for $\langle x\rangle$ and $\langle (x)^2\rangle$ can be exactly recovered from the solution of Eq. \eqref{eq_micro_P}, hence the linear growth of averaged position and variance is expected to emerge as soon as Eq. \eqref{eq_micro_P} is valid.

\end{document}